\pdfoutput=1

\documentclass[11pt]{article}

\usepackage[]{ACL2023}

\usepackage{times}
\usepackage{latexsym}

\usepackage[T1]{fontenc}

\usepackage[utf8]{inputenc}

\usepackage{microtype}

\usepackage{inconsolata}

\usepackage{colortbl}
\usepackage{booktabs}
\usepackage{ragged2e} 
\usepackage{booktabs,makecell, multirow, tabularx}

\usepackage[normalem]{ulem}
\useunder{\uline}{\ul}{}
\usepackage{threeparttable}
\usepackage{makecell}
\usepackage{tcolorbox}
\usepackage{booktabs}

\usepackage{graphicx}
\usepackage{subfigure}
\usepackage{multirow}
\usepackage{multicol}
\usepackage{pifont}
\usepackage{enumitem}
\usepackage{amsmath}
\usepackage{wasysym}
\usepackage{xspace}

\usepackage{pifont}

\renewcommand{\paragraph}[1]{\noindent\textbf{#1}.~}

\newcommand{\citea}[1]{\citeauthor{#1}~(\citeyear{#1})}

\newcommand{\increase}[1]{\textcolor{red}{\scriptsize #1}}

\title{Self-Edit: Fault-Aware Code Editor for Code Generation}

\author{Kechi Zhang, \ Zhuo Li, \ Jia Li \male, \ Ge Li\footnotemark[1], \ Zhi Jin\footnotemark[1] \\
Key Lab of High Confidence Software Technology (PKU), Ministry of Education \\
School of Computer Science, Peking University, China \\
\texttt{\{zhangkechi,lizhmq\}@pku.edu.cn}, \texttt{lijia@stu.pku.edu.cn}, \\ \texttt{\{lige,zhijin\}@pku.edu.cn}}

\begin{document}
\maketitle
\renewcommand{\thefootnote}{\fnsymbol{footnote}}
\footnotetext[1]{Corresponding authors}
\renewcommand{\thefootnote}{\arabic{footnote}}
\begin{abstract}
Large language models (LLMs) have demonstrated an impressive ability to generate codes on competitive programming tasks. However, with limited sample numbers, LLMs still suffer from poor accuracy. Inspired by the process of human programming, we propose a generate-and-edit approach named Self-Edit that utilizes execution results of the generated code from LLMs to improve the code quality on the competitive programming task. 
We execute the generated code on the example test case provided in the question and wrap execution results into a supplementary comment. Utilizing this comment as guidance, our fault-aware code editor is employed to correct errors in the generated code.
We perform extensive evaluations across two competitive programming datasets with nine different LLMs. Compared to directly generating from LLMs, our approach can improve the average of pass@1 by 89\% on APPS-dev, 31\% on APPS-test, and 48\% on HumanEval over nine popular code generation LLMs with parameter sizes ranging from 110M to 175B. 
Compared to other post-processing methods, our method demonstrates superior accuracy and efficiency.

\end{abstract}

\section{Introduction}


Large language models (LLMs) have recently been applied to the competitive programming task. This task requires understanding a complex natural language description of a problem with example test cases and correctly implementing solutions that can span hundreds of lines. Solutions are evaluated by executing them on hidden test cases. 
However, existing LLMs often have low accuracy and pass rates in this task. For example, on a popular competitive programming benchmark \textit{APPS-test} \cite{APPS}, the nearly most powerful model GPT3 \cite{gpt3} achieves only 7\% accuracy when allowed to submit only one program per task (referred to as \textit{pass@1}).

\begin{figure}[t]

\subfigure[\scriptsize]{
\label{fig:inspiration}
\includegraphics[width=\columnwidth]{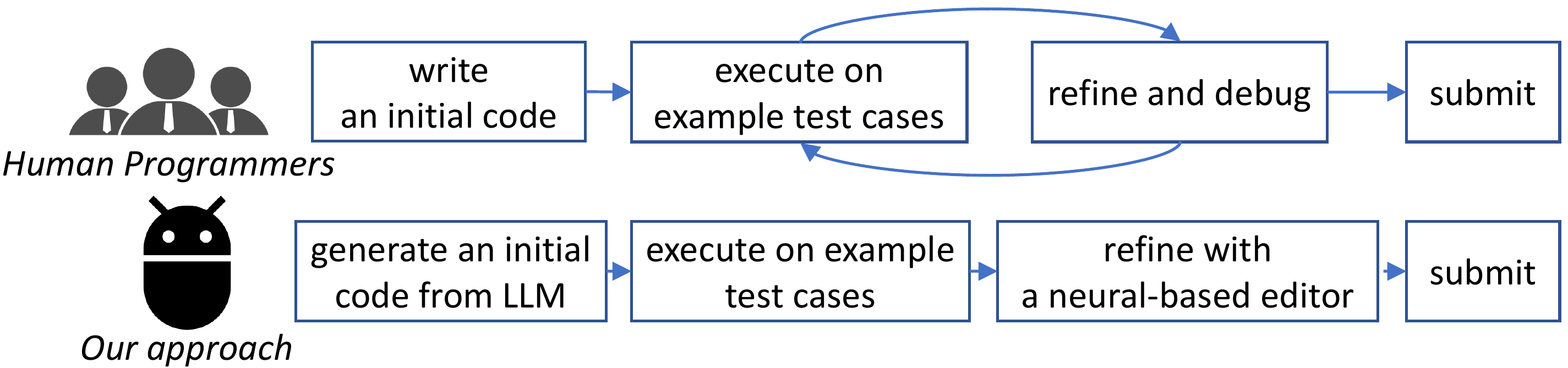}  
}
\vskip -5pt
\subfigure[\scriptsize]{
\label{fig:examplecase}
\includegraphics[width=\columnwidth]{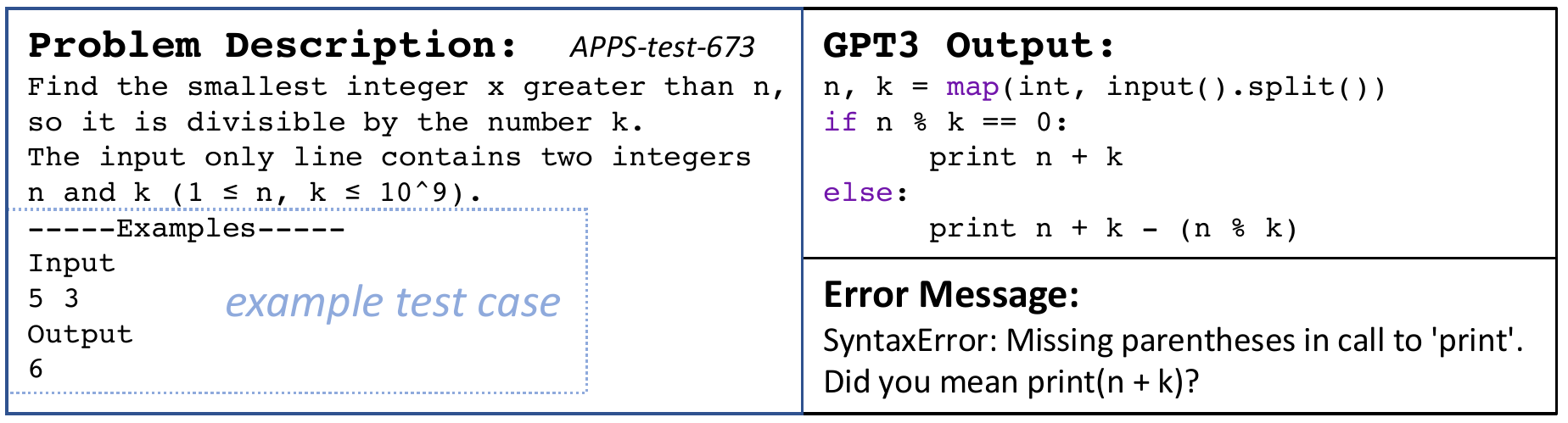}  
}
\vspace{-15pt}
\caption{(a) Our approach is inspired by the problem-solving process of human programmers. (b) Output from \textit{GPT3} model on APPS-test dataset  and its corresponding error message, which is obtained by running on the example test case.}

\end{figure}
To improve the performance of LLMs on the competitive programming task, we take inspiration from the process of human programming.
When solving competitive programming problems, programmers usually write an initial program, execute some example test cases, and refine the code based on the test results. In this process, a programmer can take key information (e.g, program outputs or compile/runtime error message) from the test results, which helps them debug the program.
We instantiate this idea by adopting a similar pipeline with a neural-based editor (in Figure \ref{fig:inspiration}). Analyzing the code generated by a pre-trained LLM, we have found that some of the generated codes can be improved with minor modifications. Figure \ref{fig:examplecase} shows an example of generated code by GPT3 on the APPS-test dataset. GPT3 generates code that is inconsistent with the problem description. We notice that the error message directly points out the bug in the code, with which we can quickly fix the error. It motivates us to investigate approaches to edit and improve the quality of the code generated by LLMs with the help of execution results.
In this work, we propose a novel generate-and-edit approach to augment LLMs on the competitive programming task, named Self-Edit. 
To mimic the above human programmers' behavior, our approach incorporates the ability of LLMs in three steps: 
\ding{182} \textit{Generation with LLMs}. We use large language models as black-box generators and generate the program based on the problem description. 
\ding{183} \textit{Execution}. Given a generated code from LLMs, we execute it on the example test case to get the execution results. We further wrap the execution results with templates as supplementary comments to include additional helpful information for editing.
\ding{184} \textit{Edit}. We develop a fault-aware neural code editor that takes the generated code and supplementary comment as input and refines the code. 
Our code editor aims to improve the quality and accuracy of code generation using LLMs. 

We conduct extensive experiments on two public competitive programming benchmarks, including APPS \cite{APPS} and HumanEval \cite{humaneval}. We apply our approach to 9 popular LLMs with parameter sizes ranging from 110M to 175B to show the universality.
Compared to directly generating from LLMs, we have several findings: \ding{182} Our approach significantly improves the performance of LLMs. 
In particular, our approach improves the average of pass@1 by 89\% on APPS-dev and 31\% on APPS-test.
Even for the chosen largest language model GPT3-175B, our relatively small editor model can improve pass@1 from 26.6\%  to 32.4\% on the APPS-dev benchmark. 
\ding{183} Our approach is generalizable on a different style of dataset HumanEval, improving the average of pass@1 by 48\%, showing the transfer ability on the out-of-distribution benchmark. 

Recently some approaches are also proposed to post-process programs generated by LLMs \cite{CodeTranslationwithExecution2022,inala2022faultaware,codet,coderreviewer}. These approaches do large-scale sampling from LLMs, rerank these sampled programs, and output the final program. 
In comparison, our self-edit framework has two advantages: 
\ding{182} Our approach maintains a constant sample budget and significantly reduces the computational overhead for LLMs. \ding{183} Our editor directly modifies the programs and outperforms these reranking-based methods, especially with a limited sample budget such as pass@1. \textbf{\textit{To our knowledge, we are the first to adopt an editing-based post-processing method for competitive programming tasks.}}

The contributions are listed as follows:
\begin{itemize}
    \item We propose a generate-and-edit approach named Self-Edit for large language models (LLMs) to generate high-quality code for competitive programming tasks.
    \item We develop a fault-aware neural code editor that takes the generated code and error messages as input and uses them to refine the code, improving its quality and accuracy.
    \item We conduct experiments on two popular datasets and nine LLMs to demonstrate the effectiveness and universality of our approach.
\end{itemize}


\section{Related Work}
\subsection{Code Generation}

Code generation is a process in which source code is automatically generated based on functional requirements
such as natural language descriptions \cite{iyer2018mapping, yin2018tranx,TiP, skcoder, acecoder} or pseudo code algorithms \cite{kulal2019spoc, oda2015django} or a old version of code \cite{codeeditor} or a response from programming tools \cite{Zhang2023ToolCoderTC}. One particularly challenging type of code generation task is competitive programming \cite{alphacode}, in which models must solve problems at the level of programming competitions. This task often involves natural language descriptions and example input-output pairs. The performance of a code generation model on competitive programming tasks can serve as a measure of its ability to create complete solutions to problems. 
In recent years, large pre-trained language models such as AlphaCode \cite{alphacode} and the GPT3 \cite{gpt3} series have demonstrated impressive capabilities in code generation and competitive programming. Other open-source code generation models include GPT-Neo \cite{gpt-neo}, GPT-J \cite{gpt-j}, CodeParrot \cite{wolf-etal-2020-transformers}, PolyCoder \cite{PolyCoder}, CodeGen \cite{codegen} and InCoder \cite{incoder}. We utilize the \textit{text-davinci-002} API from OpenAI and various competitive code generation models in this work.

\subsection{Post-processing of LLMs for code generation}

To find the correct code solutions based on LLMs, researchers adopt various post-processing methods to filter/rerank the original outputs from LLMs. In the domain of solving math problems, \citea{trainVerifier} and \citea{Generate_and_rank} chose the one with the highest rank by a trained ranker. Similar ranking methods are also used in the field of cross-domain adaptation \cite{DBLP:conf/wsdm/LiTHXCJ22}.
In the domain of code generation, post-processing techniques are also often used \cite{InteractiveCodeGeneration, coderl}.  AlphaCode \cite{alphacode} and \citea{CodeTranslationwithExecution2022} adopted the clustering and filtering methods based on the execution output of the generated programs.  \citea{inala2022faultaware} trained a fault-aware neural ranker to rerank the outputs with a large sample budget. \citea{codet} use the large models to generate test cases for themselves and automatically rank the solutions based on the test-driven dual execution agreement. \citea{coderreviewer} reranked the LLM outputs with the generation probability of back translation.

However, these existing methods require large-scale sampling. They need to generate a large number of programs for post-processing. For example, AlphaCode \cite{alphacode} needs 1 million samples per problem, costing $10^5$ TPU-seconds. In the real world, computing resources are precious and limited, and existing methods are ineffective in practical applications. Our self-edit approach addresses this issue by maintaining a constant sample budget and improving computational efficiency, described in Section \ref{sec:ranking_rst}.



\section{Methodology}

\begin{figure}[t]

\includegraphics[width=\columnwidth]{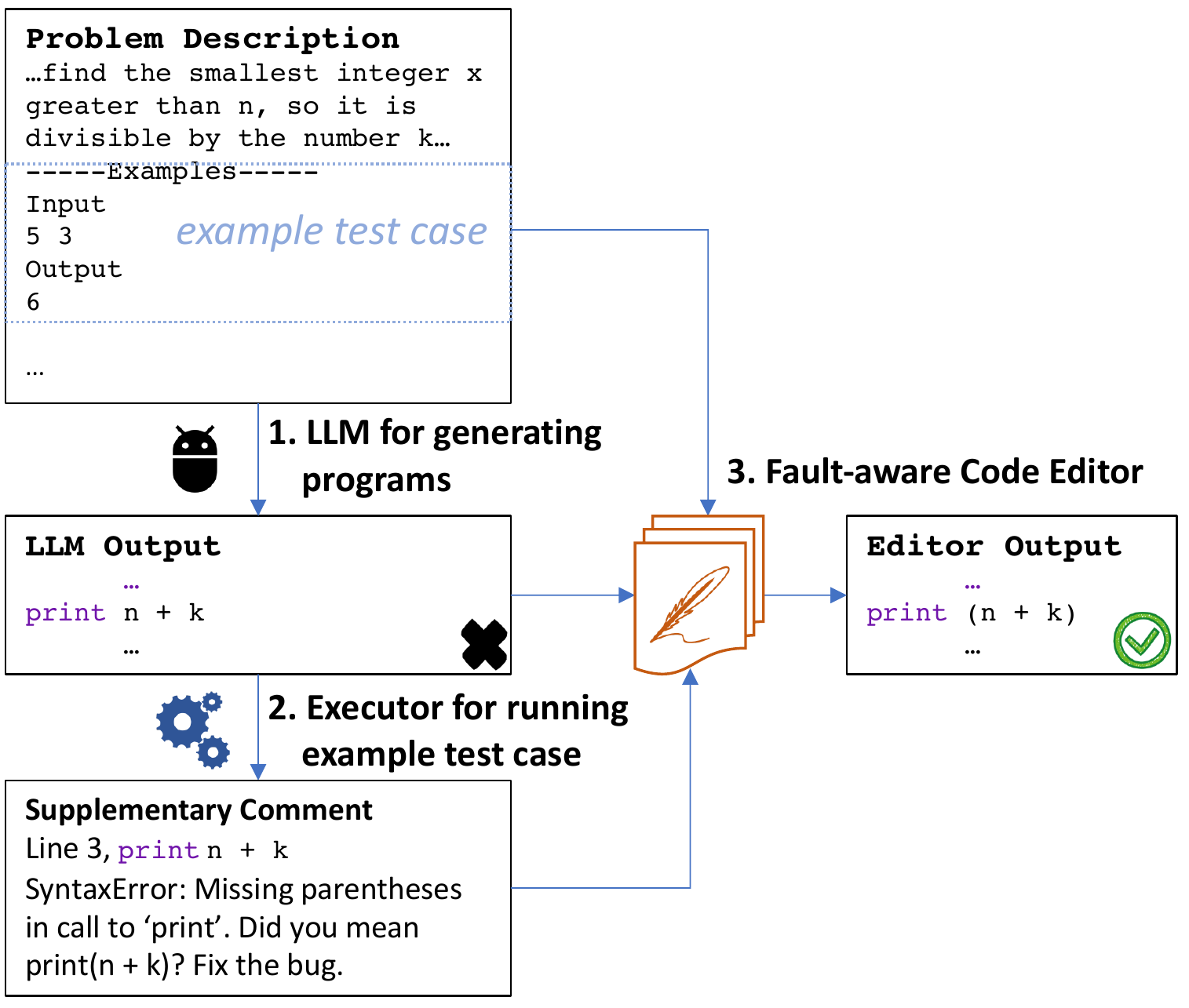}  
\caption{Pipeline of our self-edit approach.}
\label{fig:pipeline}
\end{figure}

We provide an overview of the self-edit pipeline in Figure \ref{fig:pipeline}. Given the problem description, We first generate the initial code with LLM. Then we execute the example test case to obtain test results and construct the supplementary comment.
Finally, we train a fault-aware code editor model to refine the code based on the problem description, generated code, and supplementary comment.

\subsection{LLMs as Black-box Generator}

\label{sec:LLM}
We use large language models as black-box generators with fixed parameters in our design. This design choice is motivated by the fact that training LLMs is costly, and access to LLMs is often restricted. ({\textit{E}.\textit{g}.}, OpenAI only offers paid API to infer GPT3.) Using LLM as a black-box generator makes our approach flexible for using different LLMs.
We investigate nine LLMs for code generation with sizes ranging from 110M to 175B.
A detailed comparison is described in Table \ref{tab:apps_dev}.

\subsection{Executor and Supplementary Comments}
\label{sec:comments}

After we generate the code using LLMs, we use an executor to run the example test case. We classify the execution results into three types: \ding{182} Passed: The program passes the test case. \ding{183} Wrong Answer: The program runs normally but gives incorrect outputs. \ding{184} Error: The program terminates abnormally due to syntax error, runtime exceptions, or exceeding time limit.

\begin{figure}[t]
\subfigure[\scriptsize \textit{PyCodeGPT-110M-finetuned}]{
\label{fig:pycodegpt_comment}
\includegraphics[width=0.46\columnwidth]{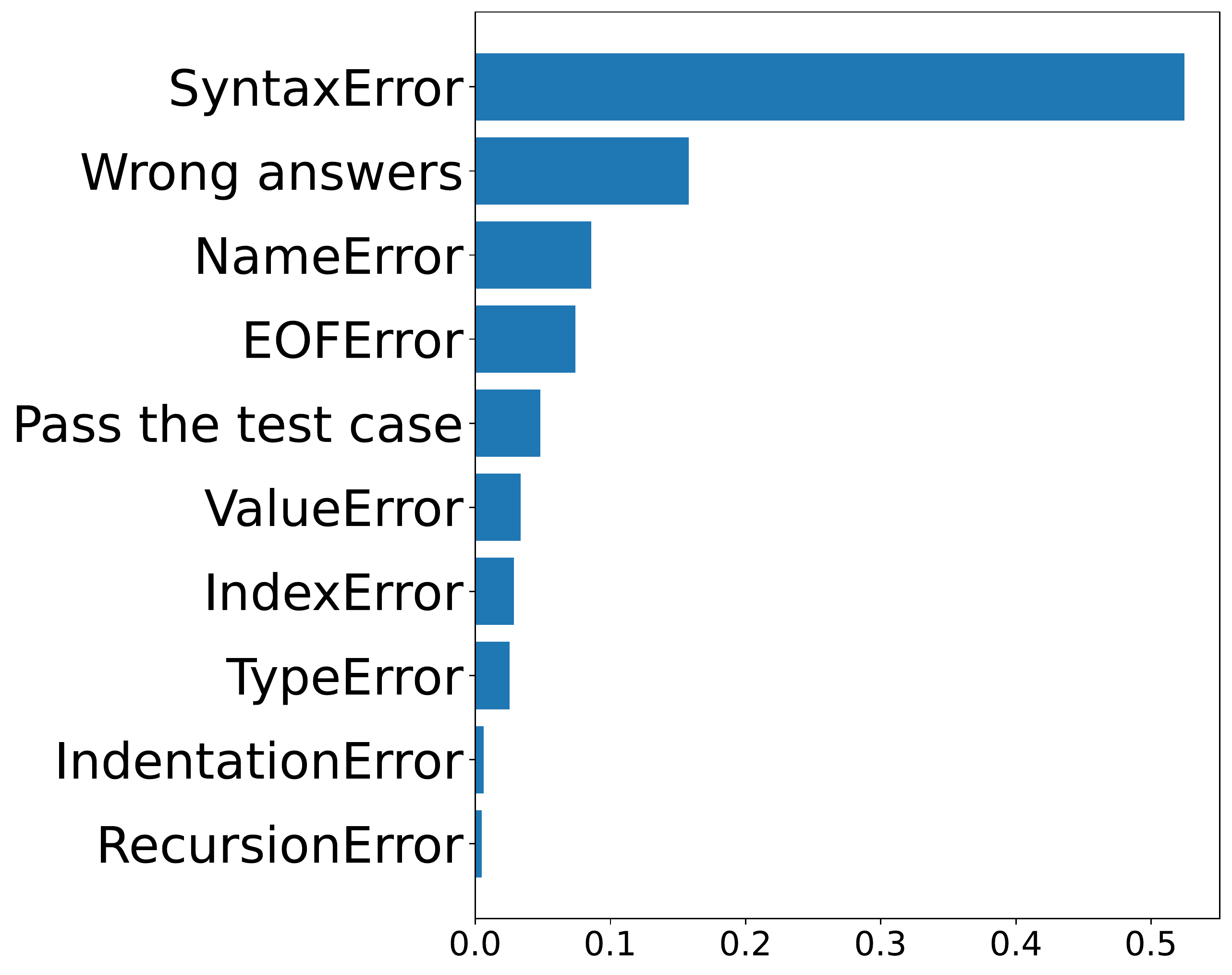}  
}
\subfigure[\scriptsize \textit{GPT3}]{
\label{fig:gpt3_comment}
\includegraphics[width=0.46\columnwidth]{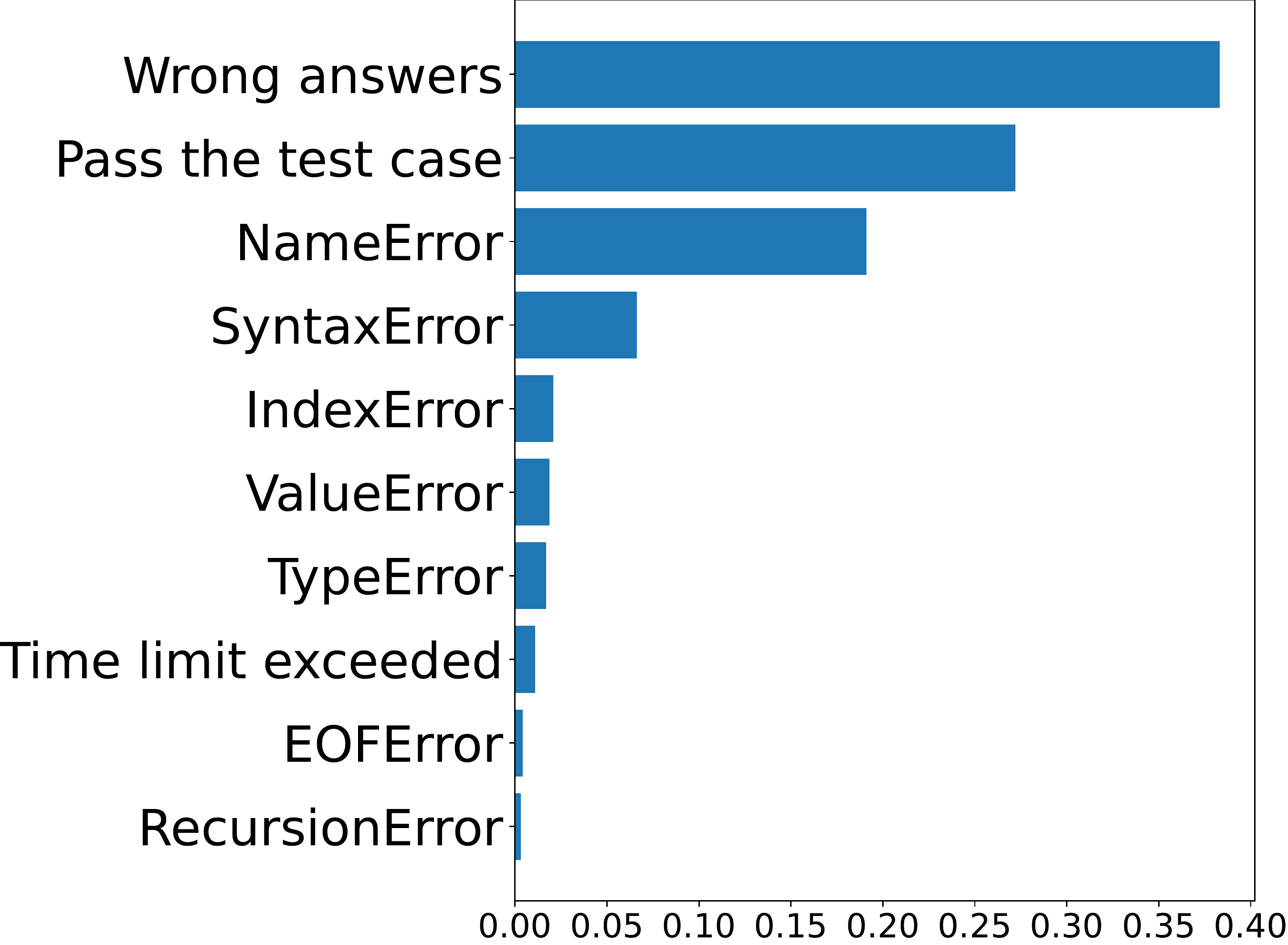}  
}
\caption{Distribution of the top 10 classes of supplementary comments in the APPS-train dataset when using the \textit{PyCodeGPT-110M-finetuned} and \textit{GPT3} models, expressed as a percentage of the total number of generated programs for each class.}
\label{fig:comment_distribution}
\end{figure}

We analyze the distribution of test results on APPS-train dataset for code generated by a relatively small model PyCodeGPT-110M and a large model GPT3-175B as shown in Figure \ref{fig:comment_distribution}. We observe that programs produced by different models yield different test result distributions.
Code generated by smaller models (PyCodeGPT) tends to encounter SyntaxError issues more frequently, while large models (GPT3) show fewer SyntaxErrors, fewer RuntimeErrors, but more normally executed cases. 

\begin{figure}[t]

\includegraphics[width=\columnwidth]{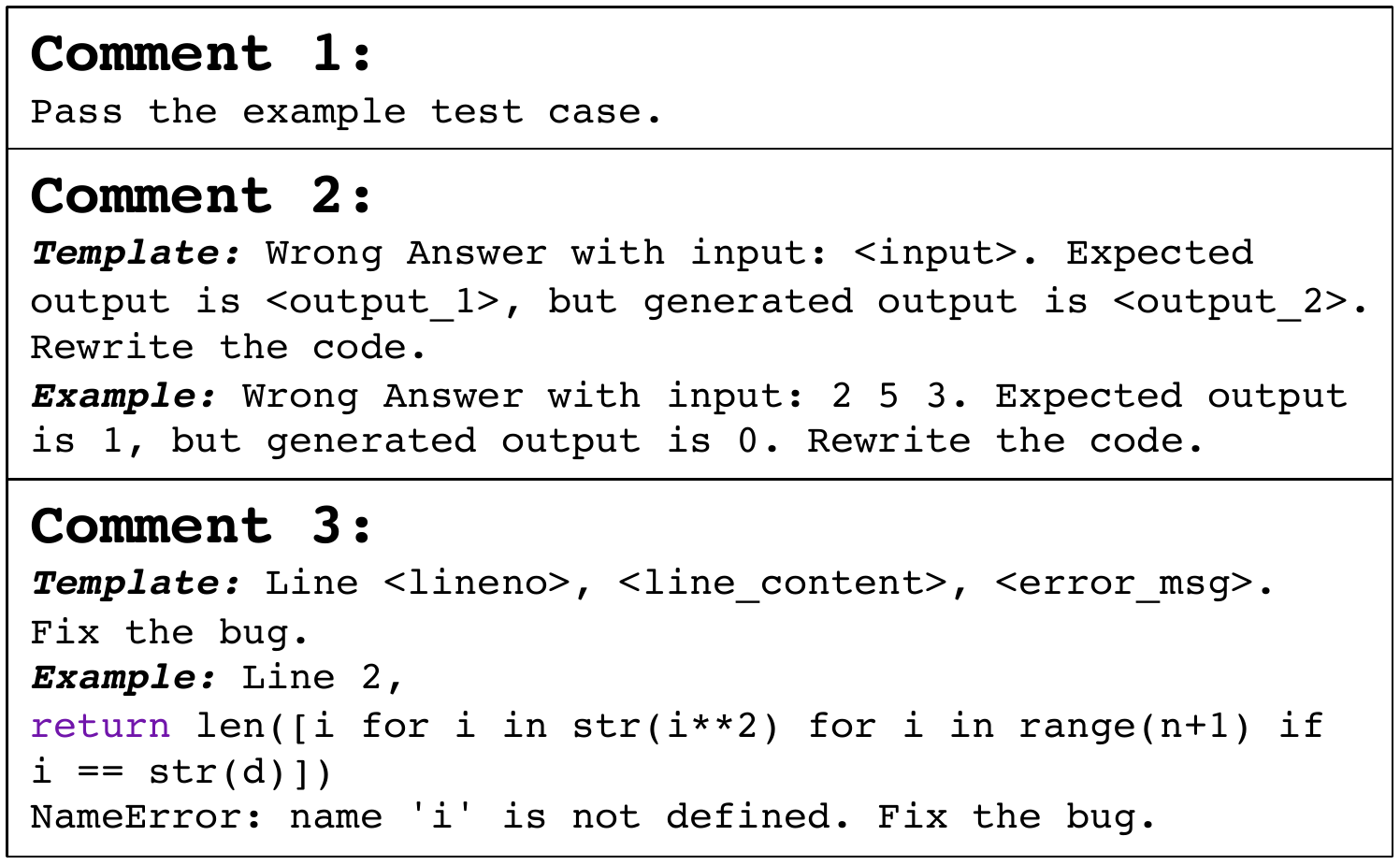}  
\caption{Example Supplementary Comments in different situations.}
\label{fig:commentexample}
\end{figure}

In order to construct meaningful supplementary comments for the code editor model to understand various execution results, we design the comment templates (Fig. \ref{fig:commentexample}) for the three types of test results. The comment template can wrap potential error messages with additional helpful information for editing. \ding{182} For the code passing the example test case, we use \textit{Comment 1}: ``Pass the example test case.''. \ding{183} For the code producing incorrect outputs, we use \textit{Comment 2} to include the relevant input, expected output, and the actual output. We also append the instruction ``Rewrite the code'' to guide the editor model to reimplement the algorithm to produce correct outputs. \ding{184} For the code that terminates with errors, we use \textit{Comment 3} to include the error line number, line context, and full error message. These supplementary comments provide additional context and clarity for the generated code and are used to guide editing the code.

\subsection{Fault-aware Code Editor}

Once we have constructed the supplementary comments, we train a fault-aware editor that takes the natural language description, generated code, and supplementary comments as input and produces higher-quality refined code.

\subsubsection{Code Editor Models}
The fault-aware code edit task is formally defined as a sequence-to-sequence task: given a natural language description $N$, a program generated by LLM $S$, and accompanied supplementary comments $C$ (Sec. \ref{sec:comments}), the model is required to generate higher-quality code $\hat{C}$ that implements the natural language description and passes test cases.
In our experiments, the input pair $(N, S, C)$ is segmented into three parts and concatenated using special separator tokens, represented as $[SOS], n_1, n_2, \dots, n_{|N|}, [CODE],s_1, \dots, s_{|S|},\\, [CMNT], c_1, \dots, c_{|C|},[EOS]$, where the lowercase letters represent the token of the corresponding content in the input pair $(N, S, C)$. We train a decoder-only model to complete the code edit task. Concretely, we implement the code editor by fine-tuning \textit{PyCodeGPT-110M} on this task.

At inference time, we first generate multiple programs from LLMs using natural language description as input. For each generated program, we feed the example test case provided in the description into the executor to obtain a fault-aware comment. We then use the editor to generate a new program, which is the final version for further evaluation. This inference approach maintains a small sample budget compared with existing large-scale sampling and filter/reranking methods.



\subsubsection{Dataset Construction for Code Editor}
To train a fault-aware code editor, we need datasets that contain the generated program and the corresponding supplementary comments. To collect such datasets, we use different LLMs (Sec. \ref{exp:setup}) to generate candidate programs for problems in the APPS-train dataset. For each problem, we sample 10 programs from the LLM and then execute the example test case to get the test results and construct supplementary comments. At this point, we get the datasets of triplets $(N, S, C)$ for different LLMs. To further obtain the ground truth program $\hat{C}$, we collect the standard ground truth programs in the original APPS training dataset and the generated programs that pass all hidden test cases. For each LLM, we create an individual editor dataset with nearly 4.5k generated programs with comments. For each generated program, we set at most 15 ground truth programs. 
As we described in Figure \ref{fig:comment_distribution}, the generated programs from different LLMs have different distributions of the corresponding comments.
To optimize the performance of the fault-aware code editor for each LLM, it is necessary to use training datasets specific to the corresponding LLM.


\subsubsection{Training Objective of Code Editor}
Editing for a high-quality program based on the input pair $(N, S, C)$ is a one-of-many task because multiple correct target programs satisfy the requirements.
Standard maximum likelihood objectives aim to minimize loss by considering all of the solutions in the training set (like recall), while we focus on a model's ability to edit a single correct solution based on the existing generated code within a limited budget of attempts (like precision). To address this discrepancy, we follow previous work and adopt a variation of GOLD \cite{GOLD, alphacode}, which incorporates an off-policy importance weight into the standard maximum likelihood objective gradient:

\begin{equation}
    \nabla\mathcal{L}(\theta )  = - \sum_{t \in \hat{C}}P_{\theta }(t) \nabla logP_{\theta}(t)
\end{equation}
where $\theta$ represents the model parameters and $logP_{\theta}(t)$ is the standard log-likelihood objective for next token prediction. The additional weight $P_{\theta}(t)$ allows the model to focus on the tokens that already have a high likelihood, so the model can concentrate on these easier-to-learn ground truth solutions and increase the chance of getting at least one correct output. Such a loss setting allows editors to learn to copy part of the content from existing generated programs to obtain better outputs. 




\section{Experiment}

We present extensive experiments that span two representative datasets and nine different LLMs for code generation, whose parameter counts range across four orders of magnitude. The details of the adopted LLMs are described in Section \ref{sec:LLM}. We aim to investigate four research questions: (1) how much can fault-aware code editors improve various code generation models on competitive programming (Sec. \ref{sec:main_rst}), (2) the advantages of editor-based methods over existing ranking methods (Sec. \ref{sec:ranking_rst}), (3) to what extent does the supplementary comments help to refine the program (Sec. \ref{sec:comment_rst}), (4) how does the number of editing rounds affect the final result (Sec. \ref{sec:numedit_rst}).

\subsection{Experiment Setup}
\label{exp:setup}

\paragraph{Dataset} We consider evaluating our approach on two existing code generation datasets: (1) \textbf{APPS} \cite{APPS}: a collection of 5000 training and 5000 test tasks collected from coding competitions and interview problems. The test set has three different difficulty levels: Introductory, Interview, and Competition.  (2) \textbf{HumanEval} \cite{humaneval}: a set of 164 test programming problems with a function signature, docstring, body, and several unit tests.
Our experiments only use the APPS-train dataset
to finetune the code generation models and the code editor models since it is the largest training dataset. Following previous studies \cite{inala2022faultaware}, we adopted the same division and used a set of 598
tasks excluded from the APPS training dataset for validation\footnote{https://github.com/microsoft/CodeRanker}.
The detailed statistic of the datasets is shown in Table \ref{tab:statistic}. The hidden test cases are those test cases for evaluation. They are not included in the problem description, so they are distinguished from the example test case used to obtain supplementary comments.

\paragraph{Base LLMs}
In this paper, we investigate the effectiveness of several widely used language models for code generation, including text-davinci-002 (175B) \cite{gpt3}, CodeGen (2B, 350M) \cite{codegen}, InCoder (1B) \cite{incoder}, GPT-Neo (1.3B, 125M) \cite{gpt-neo}, GPT-J (6B) \cite{gpt-j} and PycodeGPT (110M) \cite{CERT}. These models are evaluated under zero-shot or finetune experimental conditions, with additional descriptions provided as a part of Table \ref{tab:apps_dev}.
\footnote{
We do not use the \textit{CodeX} model as it was in closed beta and was not available during our experiments. We choose \textit{text-davinci-002} with equal parameter size as an alternative.
}

\paragraph{Editor Model}
We implement the code editor by fine-tuning \textit{PyCodeGPT-110M}. We choose this model because of its relatively small parameter size and high performance.
We also tried the \textit{CodeGen-350M} model in early experiments but found that the training speed and final performance were not as good as the model we chose.

Considering that LLMs shows strong in-context learning abilities that do not need training process, we also explore to design a variant of our self-edit method with in-context learning. We use the \textit{text-davinci-002} as both base model and editor model. The in-context learning self-edit performances are discussed in Section \ref{sec:icl}.

\begin{table}[t]
    \centering
\centering
\scriptsize
\begin{tabular}{ccccc}
\toprule
\multicolumn{1}{l}{}                                                            & \multicolumn{1}{l}{}       &              & Problems & \begin{tabular}[c]{@{}c@{}}Hidden \\ Tests\end{tabular}                          \\ \midrule
\multicolumn{1}{l}{\tiny \begin{tabular}[c]{@{}c@{}}Training \\ dataset\end{tabular}} & \multicolumn{2}{c}{APPS-train}            & 4207     & 5.56                                       \\ \midrule
\multirow{5}{*}{\tiny \begin{tabular}[c]{@{}c@{}}Testing \\ benchmark\end{tabular}}   & \multicolumn{2}{c}{APPS-dev}              & 598      & 4.03                                       \\ \cline{2-5} 
                                                                                & \multirow{3}{*}{APPS-test} & Introductory & 1000     & \multicolumn{1}{c}{\multirow{3}{*}{21.19}} \\
                                                                                &                            & Interview    & 3000     & \multicolumn{1}{c}{}                       \\
                                                                                &                            & Competition  & 1000     & \multicolumn{1}{c}{}                       \\ \cline{2-5} 
                                                                                & \multicolumn{2}{c}{HumanEval}             & 164      & 8.08       \\
\bottomrule
\end{tabular}
    \caption{Statistics of training dataset and testing benchmarks: the total number of problems in datasets (\textit{Problems}), the average number of hidden test cases per problem (\textit{Hidden Tests}). }
   \label{tab:statistic}
\end{table}

\paragraph{Metrics}
We use the metric pass rate \textit{\textbf{pass@k}} for performance evaluation and take advantage of hidden test cases to determine the functional correctness of code solutions. For each problem, we submit k code solutions for evaluation. If any of the k code solutions passes all ground truth test cases, the problem is considered solved. Then \textit{pass@k} is the percentage of solved problems.
In our experiments, we set $k = \{1, 5, 10\}$.

\begin{table*}[t]
    \centering
\centering
\small
\begin{tabular}{c|c|c
>{\columncolor[HTML]{F2F2F2}}c c
>{\columncolor[HTML]{F2F2F2}}c c
>{\columncolor[HTML]{F2F2F2}}c |c
>{\columncolor[HTML]{F2F2F2}}c}
\toprule
Code   Gen. Model               & Para. & pass@1 & \begin{tabular}[c]{@{}l@{}}edit\\ pass@1\end{tabular} & pass@5   & \begin{tabular}[c]{@{}l@{}}edit\\ pass@5\end{tabular} & pass@10  & \begin{tabular}[c]{@{}l@{}}edit\\ pass@10\end{tabular} & sol@10  & \begin{tabular}[c]{@{}l@{}}edit\\ sol@10\end{tabular}\\
\midrule
\textit{\textbf{finetuned}}      &                 &          &             &          &             &          &      & &        \\
PyCodeGPT                       & 110M            & 4.8      & 11.4        & 7.9      & 15.1        & 8.9      & 17.1  & 286 & 659       \\
GPT-Neo 125M                    & 125M            & 1.5      & 8.5    & 6.7      & 10.2    & 10.2 & 17.2 & 102 & 501     \\
CodeGen-350M                    & 350M            & 1.7      & 5.7    & 2.5      & 9.2    & 3.2      & 13.5 & 103 & 339     \\
GPT-Neo 1.3B                    & 1.3B            & 4.0      & 10.5    & 10.9      & 18.6    & 17.2      & 25.4 & 200 & 663    \\
InCoder-1B                      & 1.3B              & 9.4      & 12.4        & 12.5     & 16.2        & 13.5     & 18.1 & 568 & 730         \\
GPT-J & 6B            & 6.0     & 12.0    & 17.9      & 27.8    & 24.6      & 37.8 & 365 & 750     \\
\midrule
\textit{\textbf{zero-shot}} &                 &       &          &       &          &      &    & &       \\
InCoder-1B                      & 1.3B              & 0.2      & 4.7     & 0.8      & 7.7    & 1.2      & 9.9 & 13 & 270     \\
CodeGen-2B                      & 2.7B              & 1.3      & 7.4    & 5.9      & 14.0    & 9.7      & 19.7 & 92 & 438     \\
text-davinci-002                    & 175B            & 26.6     & 32.4        & 43.8     & 48.8        & 49.7     & 58.0    & 1626 & 1948      \\
\bottomrule
\end{tabular}
    \caption{Results on the APPS-dev dataset on how our fault-aware editors can improve the pass rates for different sample budgets with various code generation models. \textit{"finetuned"} indicates we finetune those models on APPS-train dataset. \textit{"zero-shot"} indicates we use those models in the zero-shot setting. We will use the best checkpoints of LLMs and editor models based on this validation set in other experiments.}
   \label{tab:apps_dev}
\end{table*}
\begin{table*}[t]
    \centering
\centering
\small
\begin{threeparttable}
\begin{tabular}{c|c
>{\columncolor[HTML]{F2F2F2}}c c
>{\columncolor[HTML]{F2F2F2}}c c
>{\columncolor[HTML]{F2F2F2}}c |c
>{\columncolor[HTML]{F2F2F2}}c}
\toprule
Code   Gen. Model                & pass@1   & \begin{tabular}[c]{@{}l@{}}edit\\ pass@1\end{tabular} & pass@5   & \begin{tabular}[c]{@{}l@{}}edit\\ pass@5\end{tabular} & pass@10  & \begin{tabular}[c]{@{}l@{}}edit\\ pass@10\end{tabular} & sol@10  & \begin{tabular}[c]{@{}l@{}}edit\\ sol@10\end{tabular}\\
\midrule
\textit{\textbf{finetuned}}                  &          &             &          &             &          &    & &          \\
PyCodeGPT                                   & 0.20      & 0.64        & 0.38     & 0.98        & 0.44     & 1.24 & 126 & 308        \\
GPT-Neo 125M                                & 0.08 & 0.22    & 0.40 & 0.70    & 0.70 & 1.12 & 45 & 135     \\
CodeGen 350M                                & 0.20 & 0.32    & 0.30 & 0.56    & 0.32 & 0.84 & 92 & 149     \\
GPT-Neo 1.3B                                & 0.14 & 0.68    & 0.74 & 1.38    & 1.40 & 2.10 & 106 & 340    \\
InCoder 1B                                    & 0.66     & 0.86        & 1.18     & 1.62        & 1.44     & 2.10 & 344 & 421          \\
GPT-J             & 0.70      & 1.40    & 2.46      & 3.34    & 3.52      & 4.76 & 404 & 738     \\
\midrule
\textit{\textbf{zero-shot}}                 &          &             &          &             &          &        & &      \\
InCoder 1B                                    & 0.00        & 0.24        & 0.02     & 0.50        & 0.02     & 0.76 & 1 & 121         \\
CodeGen 2B                                   & 0.12     & 0.28    & 0.34     & 0.66    & 0.66     & 1.08 & 41 & 131     \\
text-davinci-002                               & 7.48     & 7.94        & 15.94    & 16.66        & - & -  & 1876 \tnote{$\dagger$} & 1983 \tnote{$\dagger$}   \\
\bottomrule
\end{tabular}

\begin{tablenotes}
\footnotesize
     \item[$\dagger$] As we access \textit{GPT3} through a paid API, we limit the sample budget of \textit{GPT3} as 5 for this large benchmark and evaluate \textit{sol@5}.
\end{tablenotes}
\end{threeparttable}
    \caption{Results on the APPS-test dataset.}
\label{tab:apps_test}
\end{table*}

To show the number of programs corrected by our editor, we design a new metric \textit{\textbf{sol@k}}, which means the total number of correct programs given k samples per problem. For example, for the 5000 problems in APPS-test, we will generate $5000*k$ code solutions, from which we will count the number of correct solutions as \textit{sol@k}. In our experiments, we set $k = 10$.
We show the performance of the base model and the performance after editing (denoted as \textit{\textbf{edit-pass@k}} and \textit{\textbf{edit-sol@k}}).

\paragraph{Training/Inference Settings}
For each finetuned LLM, we limit the maximum epochs to 10 with a learning rate of 1e-5, and choose the best checkpoint based on the validation loss on APPS-dev. We adopt the same training strategy to train fault-aware code editors on each corresponding editor dataset. We set the maximum input length to 1024 and output length to 512 for our editors. To extract the supplementary comment, we choose only one example test case contained in the problem description even if it contains multiple. At inference time, we use temperature sampling with T = 0.8 both for LLM and editor outputs. We limit the sample budget of LLMs to 10. For each LLM output code, we only generate one code as the final version with our editor. Thus the usage of the editor maintains a constant sample budget. All experiments are conducted with 4 Tesla V100-32GB GPUs.

\subsection{Comparison with Base LLMs}
\label{sec:main_rst}
\paragraph{APPS-dev \& APPS-test}
We first compare with directly generating from LLMs to analyze how fault-aware code editors can improve nine popular code generation models. Table \ref{tab:apps_dev} shows the primary results on the APPS-dev dataset for nine different code generation models. 
The fault-aware editor improves all code generation models despite their different sizes and training settings. The average pass@1 value across nine models increases from 6.17\% to 11.67\%, representing an impressive 89\% improvement.
For those LLMs with a particularly large number of parameters, our editor can also achieve a significant improvement. For \textit{GPT3} with 175B parameters, the improvement of our editor also achieves 5.9\%, 5.0\%, 8.4\% on pass@\{1,5,10\}.

\begin{table*}[t]
    \centering
\centering
\small
\begin{tabular}{c|c
>{\columncolor[HTML]{F2F2F2}}c c
>{\columncolor[HTML]{F2F2F2}}c c
>{\columncolor[HTML]{F2F2F2}}c |c
>{\columncolor[HTML]{F2F2F2}}c}
\toprule
Code   Gen. Model               & pass@1   & \begin{tabular}[c]{@{}l@{}}edit\\ pass@1\end{tabular} & pass@5   & \begin{tabular}[c]{@{}l@{}}edit\\ pass@5\end{tabular} & pass@10  & \begin{tabular}[c]{@{}l@{}}edit\\ pass@10\end{tabular} & sol@10  & \begin{tabular}[c]{@{}l@{}}edit\\ sol@10\end{tabular}\\
\midrule
\textit{\textbf{finetuned on APPS}}                      &          &             &          &             &          &    & &          \\
PyCodeGPT                                   & 6.10      & 8.54        & 7.32     & 10.98        & 7.93     & 13.41 & 100 & 159         \\
GPT-Neo 125M                                & 0.61 & 3.05    & 3.05 & 7.32    & 6.10 & 9.76 & 21 & 76     \\
CodeGen-350M                                & 6.10 & 7.93    & 7.32 & 9.15    & 7.32 & 10.37  & 100 & 140    \\
GPT-Neo 1.3B                                & 2.44 & 5.49    & 8.54 & 10.98    & 11.59 & 14.63 & 66 & 132     \\
Incoder-1B                                   & 6.71     & 10.37        & 8.54    & 13.41        & 9.76     & 14.63  & 112 & 169   \\
GPT-J             & 7.32      & 9.76   & 17.07      & 19.51    & 25.00      & 25.61 & 133 & 183     \\
\midrule
\textit{\textbf{zero-shot}}                  &          &             &          &             &          &     & &         \\
Incoder-1B                                   & 1.22     & 3.66        & 2.44     & 7.93        & 5.49     & 10.98    & 13  &  87     \\
CodeGen-2B                                   & 14.02    & 17.07   & 29.27     & 29.88    & 34.15     & 34.15  & 226 & 255    \\
\bottomrule
\end{tabular}
    \caption{Results on the HumanEval dataset.}

   \label{tab:humaneval}
\end{table*}

\begin{table}[t]
    \centering
\centering
\small
\begin{tabular}{clll}
\toprule
\multicolumn{1}{l}{Difficulty level} & pass@1                       & pass@5                       & pass@10                      \\
\midrule
                                     & 2.10                         & 7.40                         & 10.10                         \\
\multirow{-2}{*}{Introductory}       & \cellcolor[HTML]{F2F2F2}4.90 \increase{133\%}  & \cellcolor[HTML]{F2F2F2}10.40 \increase{40.5\%} & \cellcolor[HTML]{F2F2F2}14.20 \increase{40.6\%} \\
\midrule
                                     & 0.43                         & 1.53                         & 2.37                         \\
\multirow{-2}{*}{Interview}          & \cellcolor[HTML]{F2F2F2}0.67 \increase{53.5\%}  & \cellcolor[HTML]{F2F2F2}1.97 \increase{28.1\%} & \cellcolor[HTML]{F2F2F2}3.03 \increase{28.3\%} \\
\midrule
                                     & 0.10                         & 0.30                         & 0.40                         \\
\multirow{-2}{*}{Competition}        & \cellcolor[HTML]{F2F2F2}0.10  & \cellcolor[HTML]{F2F2F2}0.40 \increase{33.3\%} & \cellcolor[HTML]{F2F2F2}0.50 \increase{25.0\%} \\
\midrule
                                     & 0.70                         & 2.46                         & 3.52                         \\
\multirow{-2}{*}{Average}            & \cellcolor[HTML]{F2F2F2}1.40 \increase{100\%} & \cellcolor[HTML]{F2F2F2}3.34 \increase{35.8\%} & \cellcolor[HTML]{F2F2F2}4.76 \increase{35.2\%} \\
\bottomrule
\end{tabular}
    \caption{Results on the APPS-test dataset with 3 difficulty levels. We use the \textit{GPTJ-6B-finetuned} as the base model. We show the base model results (the first row) and edited results (shaded row below). The numbers in red indicate the improvements of our editor.}
   \label{tab:apps_test_difficulty_gptj}
\end{table}

Results on the APPS-test dataset are shown in Table \ref{tab:apps_test}. The test problems are more challenging than APPS-dev, which we can see by the smaller pass@k numbers. 
Our editors maintain significant improvement for models of different sizes. The absolute improvement of \textit{pass@1} covers from 0.12\% to 0.7\%, showing that the editor can solve 6 to 35 more problems on this challenging benchmark. As for \textit{sol@10}, our editors can additionally correct hundreds of generated codes from LLMs. 

In some cases, we observe that the \textit{edit-pass@1} outperforms the \textit{pass@5}. It demonstrates that editing the candidate code is very sample efficient. With the editor model, the number of required programs sampled from the LLM can be reduced.

Another interesting observation is that a smaller LLM equipped with our editor can achieve comparable performance as the super large models. For example, the \textit{GPT-Neo-125M}, \textit{GPT-Neo-1.3B}, and \textit{GPT-J} are pretrained and finetuned with the same dataset. Using the editor can fill in the gaps in the parameter sizes of this series of models. The 125M pretrained model with a 110M editor can significantly outperform a 1.3B pretrained model and even outperform the 6B pretrained model in some cases. This finding can also be observed in other experiments, showing that our editor can offer a boost approximately equivalent to a tens of times pretrained model size increase.

\paragraph{On Different Difficulty-Level Problems}
Considering that the APPS-test dataset has three difficulty levels, we further analyze the improvement on problems of different difficulty in Table \ref{tab:apps_test_difficulty_gptj}.
We choose \textit{GPT-J-6B-finetuned} as the base model because it has shown promising results on this challenging benchmark and has certain representativeness. 
The editor can improve the base model on problems of all difficulty levels but has a relatively high pass rate improvement on simple \textit{"Introductory"} problems. We find that the output of LLMs is poor on very difficult problems, making it too difficult for the editor to correct these solutions. Even so, our method slightly improves the \textit{"Competition"} problems when enlarging the sample budget from 1 to 10.


\paragraph{HumanEval}
We also measure the transfer ability of our editor on HumanEval, a dataset of different styles, in Table \ref{tab:humaneval}. The HumanEval dataset requires the model to give the function body based on the function signature, comments, and example test cases. Following the executability filter in previous work \cite{coderreviewer}, in this dataset, we only edit the outputs that can not pass the example test case. We also modify the input format to be similar to the format in the APPS dataset. 
We select several representative LLMs for evaluation within our computational capabilities.
We can again see that the editor improves the performance of all code generation models on all metrics. We notice that under larger sample budget conditions, even if the pass@10 does not increase for \textit{CodeGen-2B}, our editor can still correct more generated solutions. Thus the \textit{sol@10} increases significantly.
These results demonstrate the ability and generality of our editor to correct out-of-distribution output codes.

\subsection{Comparison with Post-processing Baseline}
\label{sec:ranking_rst}
\begin{table}[t]
    \centering
\centering
\small
\begin{threeparttable}
\begin{tabular}{cccccc}
\toprule
\multicolumn{1}{l}{}           &          & \multicolumn{2}{c}{APPS-dev}  & \multicolumn{2}{c}{APPS-test} \\
\midrule
 {\scriptsize Setting}  & {\scriptsize Samples}    & @1        & @5        & @1        & @5        \\
\midrule
 {\scriptsize base model} &   & 4.0             & 10.9          & 0.14          & 0.74          \\
 \midrule
 {\scriptsize + ranker\tnote{$\dagger$}}  & 100 & 8.0             & 15.1          & 0.3           & 1.1           \\

 {\scriptsize + editor} & \{1,5\} & \textbf{10.5} & \textbf{18.6} & \textbf{0.68} & \textbf{1.38} \\
\bottomrule
\end{tabular}
\begin{tablenotes}
\footnotesize
     \item[$\dagger$] The results are copied from the original paper.
\end{tablenotes}
\end{threeparttable}
\caption{Pass Rate Results compared with CodeRanker on the APPS dataset. \textit{"+ ranker"} numbers are cited from \citea{inala2022faultaware}. We use the \textit{GPT-Neo-1.3B-finetuned} as the base model. Our method outperforms CodeRanker with an extremely small sample budget.}
\label{tab:compared_rerank}
\end{table}
This experiment compares our self-edit approach with existing post-processing methods for code generation. We choose to compare with CodeRanker \cite{inala2022faultaware}, a state-of-the-art reranking method on the APPS dataset. CodeRanker finetuned CodeBERT (125M) to classify the potential error type and use this classification prediction to rerank the generated codes from LLMs. The supervised training task makes this method more efficient than previous filtering and reranking methods. However, our experiments (Table \ref{tab:compared_rerank}) prove that our editor outperforms this state-of-the-art method in terms of accuracy and efficiency.

We choose the \textit{GPT-Neo-1.3B-finetuned} as the base model and finetune on the APPS-train dataset, keeping the same experimental settings as CodeRanker for a fair comparison.
Our method (\textit{"+ editor"}) significantly outperforms CodeRanker (\textit{"+ ranker"}). In particular, on APPS-test, our method can improve pass@1 from 0.14\% to 0.68\%, while their method can only improve from 0.14\% to 0.3\%. It means our method can solve 19 more problems on this challenging dataset. We also provide the performance of other reproduced base models in Table \ref{tab:compared_rerank_other}, where our method generally outperforms.

More importantly, existing post-processing methods rely on sampling many outputs from LLMs. For instance, the CodeRanker requires 100 outputs for each problem and then selects $k$ samples with their ranker model to evaluate \textit{pass@k} metric. In contrast, our method only requires $k=\{1,5\}$ outputs per problem and then utilizes these outputs to generate a final solution through editing. Our approach is more efficient and effective, especially when obtaining outputs from large language models is costly. As a result, our method has greater practical significance and is more suitable for use with limited sample budgets. 


\subsection{Ablation on Supplementary Comments}
\label{sec:comment_rst}
\begin{table}[t]
    \centering
\centering
\small
\begin{tabular}{l|lll|lll}
\toprule
               & \multicolumn{3}{c|}{APPS-dev}        & \multicolumn{3}{c}{APPS-test}       \\
\midrule
 Setting  & {\scriptsize @1} & {\scriptsize @5} & {\scriptsize @10}  & {\scriptsize @1} & {\scriptsize @5} & {\scriptsize @10}  \\
\midrule
 {base model}    & 4.8    & 7.9    & 8.9         & 0.2    & 0.4    & 0.4          \\
 {after edit} & 11.4   & \textbf{15.1}   & \textbf{17.1}        & \textbf{0.6}    & \textbf{1.0}    & \textbf{1.2}          \\
 \midrule
 {\ \ - comments} & 9.4 & 11.5 & 13.5  & 0.3 & 0.3 & 0.4  \\
 {\ \ + edit round}  & \textbf{11.7}    & \textbf{15.2}   & \textbf{17.1}        & 0.4    & 0.7    & 0.9         \\
 
\bottomrule
\end{tabular}
    \caption{Pass Rate Results of ablation studies. We use the \textit{PyCodeGPT-110M-finetuned} as the base model. The column \textit{"after edit"} means the performance of our editor in original settings. We experiment with additional editing rounds or without supplemental comment.}
   \label{tab:ablation}
\end{table}
To investigate the influence of supplementary comments, we remove the supplementary comments from the editor input and only use problem description and generated code to train a new editor.
Other settings are kept the same. Results on APPS validation and test datasets are shown in Table \ref{tab:ablation}.


We find that the pass rate of the modified editor decreases significantly on both datasets compared with the original editor. The modified editor can improve the APPS-dev dataset compared to the base model. However, on the more difficult APPS-test dataset, the editor model without comments shows no performance improvement. The results indicate that losing the guidance of the supplementary comment will hurt the performance of the editor model. Our experiments show that using error messages as supplementary comments for the code editor is crucial for achieving remarkable performances. 

\subsection{Ablation on the Number of Edit Rounds}
\label{sec:numedit_rst}


In our self-edit approach, we make edits to the output of LLMs to produce the final program. It leads to a question: what if we make additional edits to the program after the first edit? 
We add an additional editing step to answer this question using our original editor. Concretely, the edited program is executed on an example test case to obtain comments and then refined by the editor model again.
The results of this approach are presented in Table \ref{tab:ablation}, with the column labeled \textit{"+ edit round"} indicating the two-round editing approach.

The results show the two-round editing leads to a slight increase in pass@1 on APPS-dev. However, the additional edit round hurts the performance on APPS-test.
We guess the reason is the gap between training and test time in the second editing round. The editor is trained to edit LLM outputs but used to edit its own output in the second edit round.
In this setting, an additional editing round is not very helpful in generating better programs.

\section{Discussion}
\subsection{Time Cost compared with Post-processing Baseline}
For the specific issue of time cost, we use \textit{Google Colab} \footnote{https://colab.research.google.com} with a Tesla T4 GPU to build a demo and conduct evaluations over APPS-test dataset. We use \textit{text-davinci-002} as the base model and the average time cost is nearly 8.4s to obtain 1 sample for each question. The executor costs <0.01s, and our editor costs 3.7s to get the final output, which is acceptable in our actual experience using the demo.
By contrast, the state-of-the-art reranking method CodeRanker requires >110s to obtain candidate lists and 0.53s for the following ranker. As a result, our framework achieves better performance with less total time cost and fewer LLM calls.

\subsection{Performances of In-Context Learning Self-Edit}
\label{sec:icl}
\begin{table}[t]
    \centering
\centering
\small
\begin{tabular}{lccc|c}
\toprule
\multicolumn{1}{l}{Benchmark} & & pass@1                       & pass@5                       & sol@5                      \\
\midrule
                             & {\scriptsize before}        & 7.48                        & 15.94                         & 1876                         \\
\multirow{-2}{*}{APPS-test} & {\scriptsize after}      & \cellcolor[HTML]{F2F2F2}\textbf{8.94}   & \cellcolor[HTML]{F2F2F2}\textbf{17.12}  & \cellcolor[HTML]{F2F2F2}\textbf{2214} \\
\midrule
                           & {\scriptsize before}          & 34.76                         & 60.98                         & 288                         \\
\multirow{-2}{*}{HumanEval}  & {\scriptsize after}        & \cellcolor[HTML]{F2F2F2}\textbf{39.63}   & \cellcolor[HTML]{F2F2F2}\textbf{64.63}  & \cellcolor[HTML]{F2F2F2}\textbf{331}  \\
\bottomrule
\end{tabular}
    \caption{Results of the in-context learning self-edit on APPS-test and HumanEval benchmarks. We use the \textit{text-davinci-002} as the base model and editor model. We use the in-context learning ability of \textit{GPT3} to self-edit the model output. The constructed supplementary comments are used as input prompts for the editor.  We show the base model results (the first row) and edited results (shaded row below).}
   \label{tab:icl}
\end{table}
Given that LLMs have demonstrated strong in-context learning abilities without requiring any specific training, we leverage the capabilities of the \textit{text-davinci-002} model as both the base and editor models to develop a variant of our self-edit method that utilizes in-context learning. Specifically, we utilize in-context learning abilities of the model to self-edit its output using the supplementary comments we construct (detailed in Section \ref{sec:comments}) as input prompts for zero-shot inference. This approach allows the large model to edit its output program without additional training, offering a promising solution for optimizing the potential of LLMs.

Our experiments on APPS-test and HumanEval are presented in Table \ref{tab:icl}. Results demonstrate that our self-edit framework can be extended using in-context learning, achieving significantly better performance than smaller editors across various benchmarks. However, it is important to note that this in-context learning self-edit method still incurs a relatively large number of LLM calls. Therefore, optimizing resource requirements while exploiting the potential of LLMs remains critical. To this end, we will explore strategies to efficiently utilize the in-context learning capabilities of LLMs in our self-edit framework in future work.

\section{Conclusion}

We propose a generate-and-edit approach named Self-Edit that utilizes execution results of the generated code from LLMs to improve the code quality on the competitive programming task. The central component of our approach is the fault-aware code editor, which can edit and optimize the generated code. In-depth evaluations demonstrate our approach significantly improves the quality of LLMs' output code. 

\section{Acknowledgement}
This research is supported by the National Natural Science Foundation of China under Grant Nos. 62072007, 62192731, 62192733, 62192730, 61832009. The AI training platform supporting this work were provided by High-Flyer AI. (Hangzhou High-Flyer AI Fundamental Research Co., Ltd.) 
We also would like to thank all the anonymous reviewers for constructive comments and suggestions to this paper. 

\section*{Limitations}
Our work has several limitations, which we aim to address in our future work:

Firstly, we implement our editor with relatively small pretrained models within our computational capabilities. Our in-depth evaluations have preliminarily demostrated the effectiveness of the generate-and-edit approach. We hope to further understand the performance when using different pretrained models and architectures for the editor.

Secondly, the editor datasets we constructed are relatively small due to our computational capabilities. In our experiment, we only sample 10 programs from the LLM for each problem for dataset construction. Compared with existing post-editing methods, the dataset we use is quite small. It would be meaningful to do a detailed analysis of the impact of editor dataset size, or to experiment with other dataset construction methods. We leave this as future work.

Thirdly, We do not have strict comparison about computing resources  with other post-editing methods. In Section \ref{sec:ranking_rst} we compare with a state-of-the-art re-reaking baseline. We both use an additional model with a similar amount of parameters, but our approach outperforms using very few samples from LLMs. As accessing LLMs is costing, our approach demonstrates both superior accuracy and efficiency.

Finally, in our ablation study on the number of edit rounds, we faced with a gap between training and test time in the second editing round. Our existing implementation is not designed for this multiple-round editor. We hope to further try new specially designed model to implement the editor model.
As large language models continue to advance, the need for effective strategies to interact with LLMs will be an important area of future research.

\bibliography{reference}

\begin{thebibliography}{30}
\expandafter\ifx\csname natexlab\endcsname\relax\def\natexlab#1{#1}\fi

\bibitem[{Black et~al.(2021)Black, Gao, Wang, Leahy, and Biderman}]{gpt-neo}
Sid Black, Leo Gao, Phil Wang, Connor Leahy, and Stella Biderman. 2021.
\newblock Gpt-neo: Large scale autoregressive language modeling with
  mesh-tensorflow.
\newblock \emph{If you use this software, please cite it using these metadata},
  58.

\bibitem[{Brown et~al.(2020)Brown, Mann, Ryder, Subbiah, Kaplan, Dhariwal,
  Neelakantan, Shyam, Sastry, Askell, Agarwal, Herbert{-}Voss, Krueger,
  Henighan, Child, Ramesh, Ziegler, Wu, Winter, Hesse, Chen, Sigler, Litwin,
  Gray, Chess, Clark, Berner, McCandlish, Radford, Sutskever, and
  Amodei}]{gpt3}
Tom~B. Brown, Benjamin Mann, Nick Ryder, Melanie Subbiah, Jared Kaplan,
  Prafulla Dhariwal, Arvind Neelakantan, Pranav Shyam, Girish Sastry, Amanda
  Askell, Sandhini Agarwal, Ariel Herbert{-}Voss, Gretchen Krueger, Tom
  Henighan, Rewon Child, Aditya Ramesh, Daniel~M. Ziegler, Jeffrey Wu, Clemens
  Winter, Christopher Hesse, Mark Chen, Eric Sigler, Mateusz Litwin, Scott
  Gray, Benjamin Chess, Jack Clark, Christopher Berner, Sam McCandlish, Alec
  Radford, Ilya Sutskever, and Dario Amodei. 2020.
\newblock \href
  {https://proceedings.neurips.cc/paper/2020/hash/1457c0d6bfcb4967418bfb8ac142f64a-Abstract.html}
  {Language models are few-shot learners}.
\newblock In \emph{Advances in Neural Information Processing Systems 33: Annual
  Conference on Neural Information Processing Systems 2020, NeurIPS 2020,
  December 6-12, 2020, virtual}.

\bibitem[{Chen et~al.(2022)Chen, Zhang, Nguyen, Zan, Lin, Lou, and
  Chen}]{codet}
Bei Chen, Fengji Zhang, Anh Nguyen, Daoguang Zan, Zeqi Lin, Jian{-}Guang Lou,
  and Weizhu Chen. 2022.
\newblock \href {https://doi.org/10.48550/arXiv.2207.10397} {Codet: Code
  generation with generated tests}.
\newblock \emph{CoRR}, abs/2207.10397.

\bibitem[{Chen et~al.(2021)Chen, Tworek, Jun, Yuan, de~Oliveira~Pinto, Kaplan,
  Edwards, Burda, Joseph, Brockman, Ray, Puri, Krueger, Petrov, Khlaaf, Sastry,
  Mishkin, Chan, Gray, Ryder, Pavlov, Power, Kaiser, Bavarian, Winter, Tillet,
  Such, Cummings, Plappert, Chantzis, Barnes, Herbert{-}Voss, Guss, Nichol,
  Paino, Tezak, Tang, Babuschkin, Balaji, Jain, Saunders, Hesse, Carr, Leike,
  Achiam, Misra, Morikawa, Radford, Knight, Brundage, Murati, Mayer, Welinder,
  McGrew, Amodei, McCandlish, Sutskever, and Zaremba}]{humaneval}
Mark Chen, Jerry Tworek, Heewoo Jun, Qiming Yuan, Henrique~Ponde
  de~Oliveira~Pinto, Jared Kaplan, Harrison Edwards, Yuri Burda, Nicholas
  Joseph, Greg Brockman, Alex Ray, Raul Puri, Gretchen Krueger, Michael Petrov,
  Heidy Khlaaf, Girish Sastry, Pamela Mishkin, Brooke Chan, Scott Gray, Nick
  Ryder, Mikhail Pavlov, Alethea Power, Lukasz Kaiser, Mohammad Bavarian,
  Clemens Winter, Philippe Tillet, Felipe~Petroski Such, Dave Cummings,
  Matthias Plappert, Fotios Chantzis, Elizabeth Barnes, Ariel Herbert{-}Voss,
  William~Hebgen Guss, Alex Nichol, Alex Paino, Nikolas Tezak, Jie Tang, Igor
  Babuschkin, Suchir Balaji, Shantanu Jain, William Saunders, Christopher
  Hesse, Andrew~N. Carr, Jan Leike, Joshua Achiam, Vedant Misra, Evan Morikawa,
  Alec Radford, Matthew Knight, Miles Brundage, Mira Murati, Katie Mayer, Peter
  Welinder, Bob McGrew, Dario Amodei, Sam McCandlish, Ilya Sutskever, and
  Wojciech Zaremba. 2021.
\newblock \href {http://arxiv.org/abs/2107.03374} {Evaluating large language
  models trained on code}.
\newblock \emph{CoRR}, abs/2107.03374.

\bibitem[{Cobbe et~al.(2021)Cobbe, Kosaraju, Bavarian, Hilton, Nakano, Hesse,
  and Schulman}]{trainVerifier}
Karl Cobbe, Vineet Kosaraju, Mohammad Bavarian, Jacob Hilton, Reiichiro Nakano,
  Christopher Hesse, and John Schulman. 2021.
\newblock \href {http://arxiv.org/abs/2110.14168} {Training verifiers to solve
  math word problems}.
\newblock \emph{CoRR}, abs/2110.14168.

\bibitem[{Fried et~al.(2022)Fried, Aghajanyan, Lin, Wang, Wallace, Shi, Zhong,
  Yih, Zettlemoyer, and Lewis}]{incoder}
Daniel Fried, Armen Aghajanyan, Jessy Lin, Sida Wang, Eric Wallace, Freda Shi,
  Ruiqi Zhong, Wen{-}tau Yih, Luke Zettlemoyer, and Mike Lewis. 2022.
\newblock \href {https://doi.org/10.48550/arXiv.2204.05999} {Incoder: {A}
  generative model for code infilling and synthesis}.
\newblock \emph{CoRR}, abs/2204.05999.

\bibitem[{Hendrycks et~al.(2021)Hendrycks, Basart, Kadavath, Mazeika, Arora,
  Guo, Burns, Puranik, He, Song, and Steinhardt}]{APPS}
Dan Hendrycks, Steven Basart, Saurav Kadavath, Mantas Mazeika, Akul Arora,
  Ethan Guo, Collin Burns, Samir Puranik, Horace He, Dawn Song, and Jacob
  Steinhardt. 2021.
\newblock \href
  {https://datasets-benchmarks-proceedings.neurips.cc/paper/2021/hash/c24cd76e1ce41366a4bbe8a49b02a028-Abstract-round2.html}
  {Measuring coding challenge competence with {APPS}}.
\newblock In \emph{Proceedings of the Neural Information Processing Systems
  Track on Datasets and Benchmarks 1, NeurIPS Datasets and Benchmarks 2021,
  December 2021, virtual}.

\bibitem[{Inala et~al.(2022)Inala, Wang, Yang, Codas, Encarnaci{\'o}n, Lahiri,
  Musuvathi, and Gao}]{inala2022faultaware}
Jeevana~Priya Inala, Chenglong Wang, Mei Yang, Andres Codas, Mark
  Encarnaci{\'o}n, Shuvendu~K Lahiri, Madanlal Musuvathi, and Jianfeng Gao.
  2022.
\newblock \href {https://openreview.net/forum?id=LtJMqnbslJe} {Fault-aware
  neural code rankers}.
\newblock In \emph{Advances in Neural Information Processing Systems}.

\bibitem[{Iyer et~al.(2018)Iyer, Konstas, Cheung, and
  Zettlemoyer}]{iyer2018mapping}
Srinivasan Iyer, Ioannis Konstas, Alvin Cheung, and Luke Zettlemoyer. 2018.
\newblock Mapping language to code in programmatic context.
\newblock \emph{arXiv preprint arXiv:1808.09588}.

\bibitem[{Kulal et~al.(2019)Kulal, Pasupat, Chandra, Lee, Padon, Aiken, and
  Liang}]{kulal2019spoc}
Sumith Kulal, Panupong Pasupat, Kartik Chandra, Mina Lee, Oded Padon, Alex
  Aiken, and Percy~S Liang. 2019.
\newblock Spoc: Search-based pseudocode to code.
\newblock \emph{Advances in Neural Information Processing Systems}, 32.

\bibitem[{Lahiri et~al.(2022)Lahiri, Naik, Sakkas, Choudhury, von Veh,
  Musuvathi, Inala, Wang, and Gao}]{InteractiveCodeGeneration}
Shuvendu~K. Lahiri, Aaditya Naik, Georgios Sakkas, Piali Choudhury, Curtis von
  Veh, Madanlal Musuvathi, Jeevana~Priya Inala, Chenglong Wang, and Jianfeng
  Gao. 2022.
\newblock \href {https://doi.org/10.48550/arXiv.2208.05950} {Interactive code
  generation via test-driven user-intent formalization}.
\newblock \emph{CoRR}, abs/2208.05950.

\bibitem[{Le et~al.(2022)Le, Wang, Gotmare, Savarese, and Hoi}]{coderl}
Hung Le, Yue Wang, Akhilesh~Deepak Gotmare, Silvio Savarese, and
  Steven~Chu{-}Hong Hoi. 2022.
\newblock \href
  {http://papers.nips.cc/paper\_files/paper/2022/hash/8636419dea1aa9fbd25fc4248e702da4-Abstract-Conference.html}
  {Coderl: Mastering code generation through pretrained models and deep
  reinforcement learning}.
\newblock In \emph{NeurIPS}.

\bibitem[{Li et~al.(2023{\natexlab{a}})Li, Li, Li, and Jin}]{TiP}
Jia Li, Ge~Li, Yongmin Li, and Zhi Jin. 2023{\natexlab{a}}.
\newblock Enabling programming thinking in large language models toward code
  generation.
\newblock \emph{arXiv preprint arXiv:2305.06599}.

\bibitem[{Li et~al.(2022{\natexlab{a}})Li, Li, Li, Jin, Hu, Zhang, and
  Fu}]{codeeditor}
Jia Li, Ge~Li, Zhuo Li, Zhi Jin, Xing Hu, Kechi Zhang, and Zhiyi Fu.
  2022{\natexlab{a}}.
\newblock Codeeditor: Learning to edit source code with pre-trained models.
\newblock \emph{arXiv preprint arXiv:2210.17040}.

\bibitem[{Li et~al.(2023{\natexlab{b}})Li, Li, Li, Jin, Hao, and Hu}]{skcoder}
Jia Li, Yongmin Li, Ge~Li, Zhi Jin, Yiyang Hao, and Xing Hu.
  2023{\natexlab{b}}.
\newblock Skcoder: A sketch-based approach for automatic code generation.
\newblock \emph{arXiv preprint arXiv:2302.06144}.

\bibitem[{Li et~al.(2022{\natexlab{b}})Li, Tao, Hu, Xu, Chen, and
  Jiang}]{DBLP:conf/wsdm/LiTHXCJ22}
Jia Li, Chongyang Tao, Huang Hu, Can Xu, Yining Chen, and Daxin Jiang.
  2022{\natexlab{b}}.
\newblock \href {https://doi.org/10.1145/3488560.3498404} {Unsupervised
  cross-domain adaptation for response selection using self-supervised and
  adversarial training}.
\newblock In \emph{{WSDM} '22: The Fifteenth {ACM} International Conference on
  Web Search and Data Mining, Virtual Event / Tempe, AZ, USA, February 21 - 25,
  2022}, pages 562--570. {ACM}.

\bibitem[{Li et~al.(2023{\natexlab{c}})Li, Zhao, Li, Li, and Jin}]{acecoder}
Jia Li, Yunfei Zhao, Yongmin Li, Ge~Li, and Zhi Jin. 2023{\natexlab{c}}.
\newblock Towards enhancing in-context learning for code generation.
\newblock \emph{arXiv preprint arXiv:2303.17780}.

\bibitem[{Li et~al.(2022{\natexlab{c}})Li, Choi, Chung, Kushman, Schrittwieser,
  Leblond, Eccles, Keeling, Gimeno, Lago, Hubert, Choy, de~Masson~d'Autume,
  Babuschkin, Chen, Huang, Welbl, Gowal, Cherepanov, Molloy, Mankowitz, Robson,
  Kohli, de~Freitas, Kavukcuoglu, and Vinyals}]{alphacode}
Yujia Li, David~H. Choi, Junyoung Chung, Nate Kushman, Julian Schrittwieser,
  R{\'{e}}mi Leblond, Tom Eccles, James Keeling, Felix Gimeno, Agustin~Dal
  Lago, Thomas Hubert, Peter Choy, Cyprien de~Masson~d'Autume, Igor Babuschkin,
  Xinyun Chen, Po{-}Sen Huang, Johannes Welbl, Sven Gowal, Alexey Cherepanov,
  James Molloy, Daniel~J. Mankowitz, Esme~Sutherland Robson, Pushmeet Kohli,
  Nando de~Freitas, Koray Kavukcuoglu, and Oriol Vinyals. 2022{\natexlab{c}}.
\newblock \href {https://doi.org/10.48550/arXiv.2203.07814} {Competition-level
  code generation with alphacode}.
\newblock \emph{CoRR}, abs/2203.07814.

\bibitem[{Nijkamp et~al.(2022)Nijkamp, Pang, Hayashi, Tu, Wang, Zhou, Savarese,
  and Xiong}]{codegen}
Erik Nijkamp, Bo~Pang, Hiroaki Hayashi, Lifu Tu, Huan Wang, Yingbo Zhou, Silvio
  Savarese, and Caiming Xiong. 2022.
\newblock \href {https://doi.org/10.48550/arXiv.2203.13474} {A conversational
  paradigm for program synthesis}.
\newblock \emph{CoRR}, abs/2203.13474.

\bibitem[{Oda et~al.(2015)Oda, Fudaba, Neubig, Hata, Sakti, Toda, and
  Nakamura}]{oda2015django}
Yusuke Oda, Hiroyuki Fudaba, Graham Neubig, Hideaki Hata, Sakriani Sakti,
  Tomoki Toda, and Satoshi Nakamura. 2015.
\newblock Learning to generate pseudo-code from source code using statistical
  machine translation.
\newblock In \emph{2015 30th IEEE/ACM International Conference on Automated
  Software Engineering (ASE)}, pages 574--584. IEEE.

\bibitem[{Pang and He(2021)}]{GOLD}
Richard~Yuanzhe Pang and He~He. 2021.
\newblock \href {https://openreview.net/forum?id=RovX-uQ1Hua} {Text generation
  by learning from demonstrations}.
\newblock In \emph{9th International Conference on Learning Representations,
  {ICLR} 2021, Virtual Event, Austria, May 3-7, 2021}. OpenReview.net.

\bibitem[{Shen et~al.(2021)Shen, Yin, Li, Shang, Jiang, Zhang, and
  Liu}]{Generate_and_rank}
Jianhao Shen, Yichun Yin, Lin Li, Lifeng Shang, Xin Jiang, Ming Zhang, and Qun
  Liu. 2021.
\newblock \href {https://doi.org/10.18653/v1/2021.findings-emnlp.195} {Generate
  {\&} rank: {A} multi-task framework for math word problems}.
\newblock In \emph{Findings of the Association for Computational Linguistics:
  {EMNLP} 2021, Virtual Event / Punta Cana, Dominican Republic, 16-20 November,
  2021}, pages 2269--2279. Association for Computational Linguistics.

\bibitem[{Shi et~al.(2022)Shi, Fried, Ghazvininejad, Zettlemoyer, and
  Wang}]{CodeTranslationwithExecution2022}
Freda Shi, Daniel Fried, Marjan Ghazvininejad, Luke Zettlemoyer, and Sida~I.
  Wang. 2022.
\newblock \href {https://doi.org/10.48550/arXiv.2204.11454} {Natural language
  to code translation with execution}.
\newblock \emph{CoRR}, abs/2204.11454.

\bibitem[{Wang and Komatsuzaki(2021)}]{gpt-j}
Ben Wang and Aran Komatsuzaki. 2021.
\newblock {GPT-J-6B: A 6 Billion Parameter Autoregressive Language Model}.
\newblock \url{https://github.com/kingoflolz/mesh-transformer-jax}.

\bibitem[{Wolf et~al.(2020)Wolf, Debut, Sanh, Chaumond, Delangue, Moi, Cistac,
  Rault, Louf, Funtowicz, Davison, Shleifer, von Platen, Ma, Jernite, Plu, Xu,
  Scao, Gugger, Drame, Lhoest, and Rush}]{wolf-etal-2020-transformers}
Thomas Wolf, Lysandre Debut, Victor Sanh, Julien Chaumond, Clement Delangue,
  Anthony Moi, Pierric Cistac, Tim Rault, Rémi Louf, Morgan Funtowicz, Joe
  Davison, Sam Shleifer, Patrick von Platen, Clara Ma, Yacine Jernite, Julien
  Plu, Canwen Xu, Teven~Le Scao, Sylvain Gugger, Mariama Drame, Quentin Lhoest,
  and Alexander~M. Rush. 2020.
\newblock \href {https://www.aclweb.org/anthology/2020.emnlp-demos.6}
  {Transformers: State-of-the-art natural language processing}.
\newblock In \emph{Proceedings of the 2020 Conference on Empirical Methods in
  Natural Language Processing: System Demonstrations}, pages 38--45, Online.
  Association for Computational Linguistics.

\bibitem[{Xu et~al.(2022)Xu, Alon, Neubig, and Hellendoorn}]{PolyCoder}
Frank~F. Xu, Uri Alon, Graham Neubig, and Vincent~Josua Hellendoorn. 2022.
\newblock \href {https://doi.org/10.1145/3520312.3534862} {A systematic
  evaluation of large language models of code}.
\newblock In \emph{MAPS@PLDI 2022: 6th {ACM} {SIGPLAN} International Symposium
  on Machine Programming, San Diego, CA, USA, 13 June 2022}, pages 1--10.
  {ACM}.

\bibitem[{Yin and Neubig(2018)}]{yin2018tranx}
Pengcheng Yin and Graham Neubig. 2018.
\newblock Tranx: A transition-based neural abstract syntax parser for semantic
  parsing and code generation.
\newblock \emph{arXiv preprint arXiv:1810.02720}.

\bibitem[{Zan et~al.(2022)Zan, Chen, Yang, Lin, Kim, Guan, Wang, Chen, and
  Lou}]{CERT}
Daoguang Zan, Bei Chen, Dejian Yang, Zeqi Lin, Minsu Kim, Bei Guan, Yongji
  Wang, Weizhu Chen, and Jian-Guang Lou. 2022.
\newblock {CERT}: Continual pre-training on sketches for library-oriented code
  generation.
\newblock In \emph{The 2022 International Joint Conference on Artificial
  Intelligence}.

\bibitem[{Zhang et~al.(2023)Zhang, Li, Li, Li, and Jin}]{Zhang2023ToolCoderTC}
Kechi Zhang, Ge~Li, Jia Li, Zhuo Li, and Zhi Jin. 2023.
\newblock Toolcoder: Teach code generation models to use api search tools.
\newblock \emph{ArXiv}, abs/2305.04032.

\bibitem[{Zhang et~al.(2022)Zhang, Yu, Hashimoto, Lewis, Yih, Fried, and
  Wang}]{coderreviewer}
Tianyi Zhang, Tao Yu, Tatsunori~B. Hashimoto, Mike Lewis, Wen{-}tau Yih, Daniel
  Fried, and Sida~I. Wang. 2022.
\newblock \href {https://doi.org/10.48550/arXiv.2211.16490} {Coder reviewer
  reranking for code generation}.
\newblock \emph{CoRR}, abs/2211.16490.

\end{thebibliography}
\bibliographystyle{acl_natbib}

\appendix


\section{Compared with CodeRanker}
\label{sec:appendix}
\begin{table*}[t]
    \centering
\centering
\small
\begin{threeparttable}
\begin{tabular}{crccccc}
\toprule
\multicolumn{7}{l}{\textit{GPT-Neo-125M-finetuned}}                                                                          \\
\multicolumn{1}{l}{}         &     &         & \multicolumn{2}{c}{APPS-dev}  & \multicolumn{2}{c}{APPS-test} \\
\midrule
\multicolumn{1}{l}{}         & { Setting}  & Samples    & @1        & @5        & @1        & @5        \\
\midrule
\multirow{2}{*}{ \begin{tabular}[c]{@{}c@{}}Reported in\\ \cite{inala2022faultaware}\end{tabular}}     & { base model \tnote{$\dagger$}} &   & 1.4           & 5.2           & 0.04          & 0.17          \\
                             & { + ranker} & 100 & 6.5             & \textbf{11.4}          & 0.1           & 0.5           \\
\midrule
\multirow{2}{*}{ {Our results}} & { base model} &  & 1.5             & 6.7          & 0.08          & 0.40          \\
                             & { + editor} & \textbf{\{1,5\}} & \textbf{8.5} & 10.2 & \textbf{0.22} & \textbf{0.70} \\
\midrule
\midrule

\multicolumn{7}{l}{\textit{GPT-Neo-1.3B-finetuned}}    \\
\multicolumn{1}{l}{}         & &    & \multicolumn{2}{c}{APPS-dev}  & \multicolumn{2}{c}{APPS-test} \\
\midrule
\multicolumn{1}{l}{}         & { Setting} & Samples     & @1        & @5        & @1        & @5        \\
\midrule
\multirow{2}{*}{ \begin{tabular}[c]{@{}c@{}}Reported in\\ \cite{inala2022faultaware}\end{tabular}}     & { base model \tnote{$\dagger$}} &   & 2.6           & 9.1           & 0.14          & 0.53          \\
                             & { + ranker} & 100 & 8.0             & 15.1          & 0.3           & 1.1           \\
\midrule
\multirow{2}{*}{ {Our results}} & { base model} &  & 4.0             & 10.9          & 0.14          & 0.74          \\
                             & { + editor} & \textbf{\{1,5\}} & \textbf{10.5} & \textbf{18.6} & \textbf{0.68} & \textbf{1.38} \\

\midrule
\midrule
\multicolumn{7}{l}{\textit{GPT-J-6B-finetuned}}                                                                          \\
\multicolumn{1}{l}{}         & &              & \multicolumn{2}{c}{APPS-dev}  & \multicolumn{2}{c}{APPS-test} \\
\midrule
\multicolumn{1}{l}{}         & { Setting} & Samples     & @1        & @5        & @1        & @5        \\
\midrule
\multirow{2}{*}{ \begin{tabular}[c]{@{}c@{}}Reported in\\ \cite{inala2022faultaware}\end{tabular}}     & { base model \tnote{$\dagger$}} &  & 5.1           & 15.6           & 0.5          & 1.6          \\
                             & { + ranker} & 100 & 11.0             & 21.7          & 0.8           & 2.6           \\
\midrule
\multirow{2}{*}{ {Our results}} & { base model} &   & 6.0             & 17.9          & 0.7          & 2.46          \\
                             & { + editor} &\textbf{\{1,5\}} & \textbf{12.0} & \textbf{27.8} & \textbf{1.4} & \textbf{3.34} \\
\bottomrule
\end{tabular}
\begin{tablenotes}

     \item[$\dagger$] As CodeRanker does not release the weights of base models, we cite their results from \citea{inala2022faultaware} and reproduce finetuned base models shown in the \textit{"Our results - base model"} row below.
\end{tablenotes}
\end{threeparttable}
    \caption{Full details of Pass Rate Results compared with the CodeRanker on the APPS dataset. We use \textit{GPT-Neo-125M-finetuned}, \textit{GPT-Neo-1.3B-finetune} and \textit{GPT-J-6B-finetuned} as the base model.}
\vspace{-2mm}
   \label{tab:compared_rerank_other}
\end{table*}
We compare with CodeRanker \cite{inala2022faultaware} using \textit{GPT-Neo-125M-finetuned}, \textit{GPT-Neo-1.3B-finetuned} and \textit{GPT-J-6B-finetuned}  as the base model.
For fair comparison, we choose the same base model, training dataset and test benchmark as the CodeRanker. We choose the above three base models and finetune on the APPS-train dataset to reproduce their results. The purpose of this step is to make our base model results similar to their reported base model results, so as to fairly compare the post-processing performance. In the experiments, the base model performance in our results is similar to the base model reported by CodeRanker.
Full details of results are shown in Table \ref{tab:compared_rerank_other}. With a very small number of samples output by LLMs, our method significantly exceeds this state-of-the-art baseline.

\section{Qualitative analysis of Code Editor}

\begin{figure*}[t]
\subfigure[\scriptsize]{
\includegraphics[width=\columnwidth]{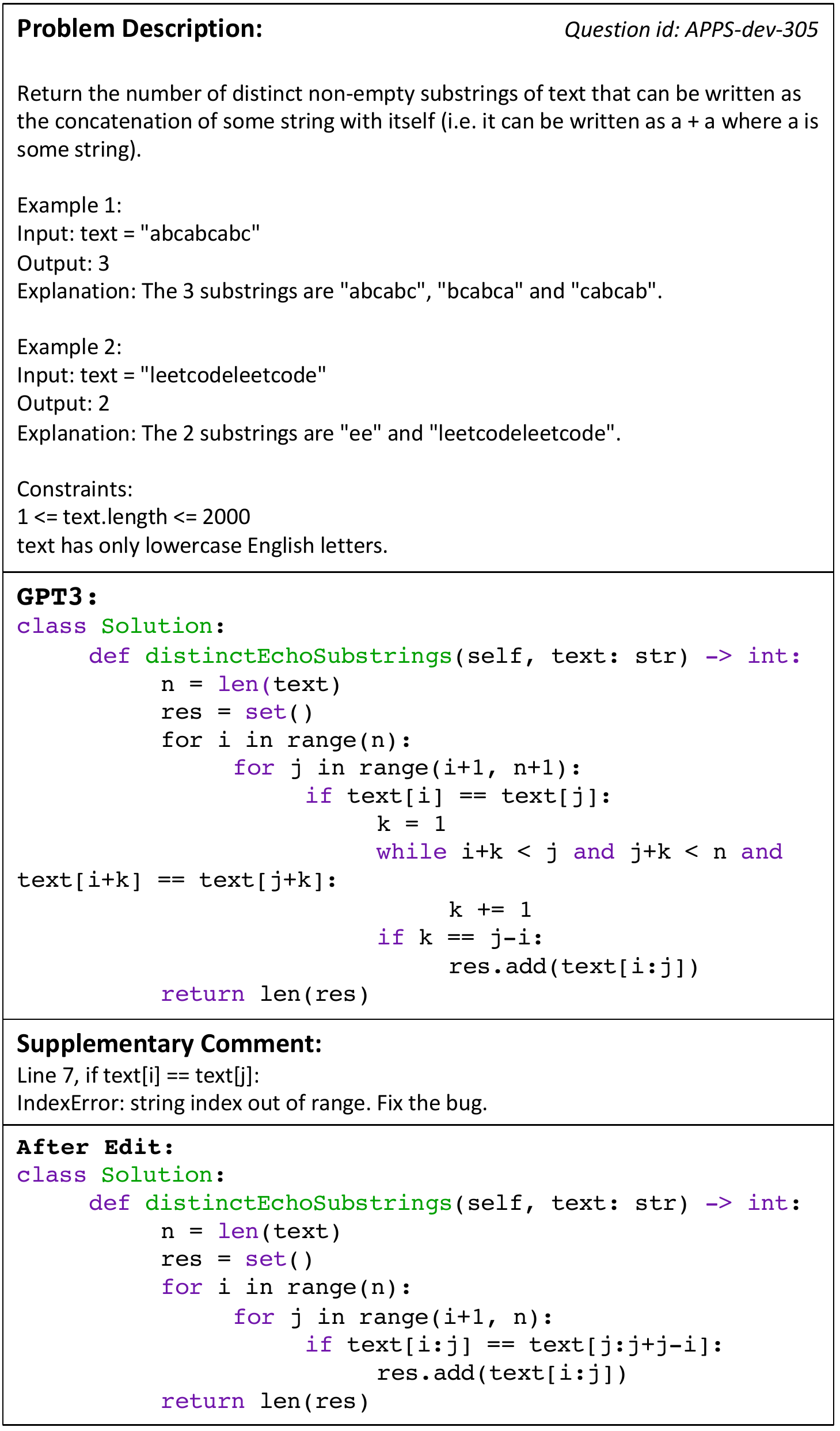}  
}
\subfigure[\scriptsize]{
\includegraphics[width=\columnwidth]{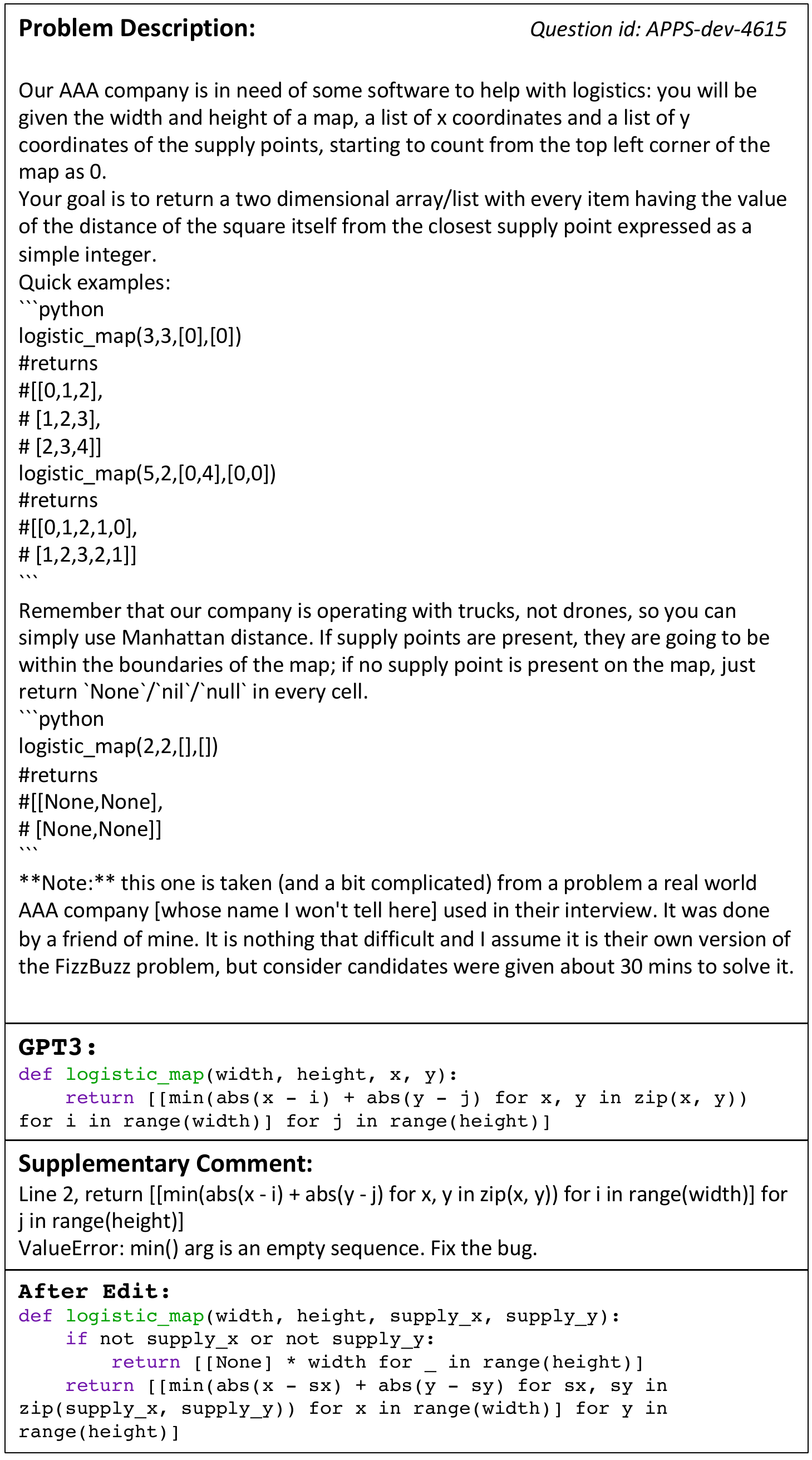}  
}
\caption{Case Study on APPS-dev dataset using \textit{GPT3} model.}
\label{fig:case_study_dev}
\vspace{-10pt}
\end{figure*}

\begin{figure*}[t]
\subfigure[\scriptsize]{
\includegraphics[width=\columnwidth]{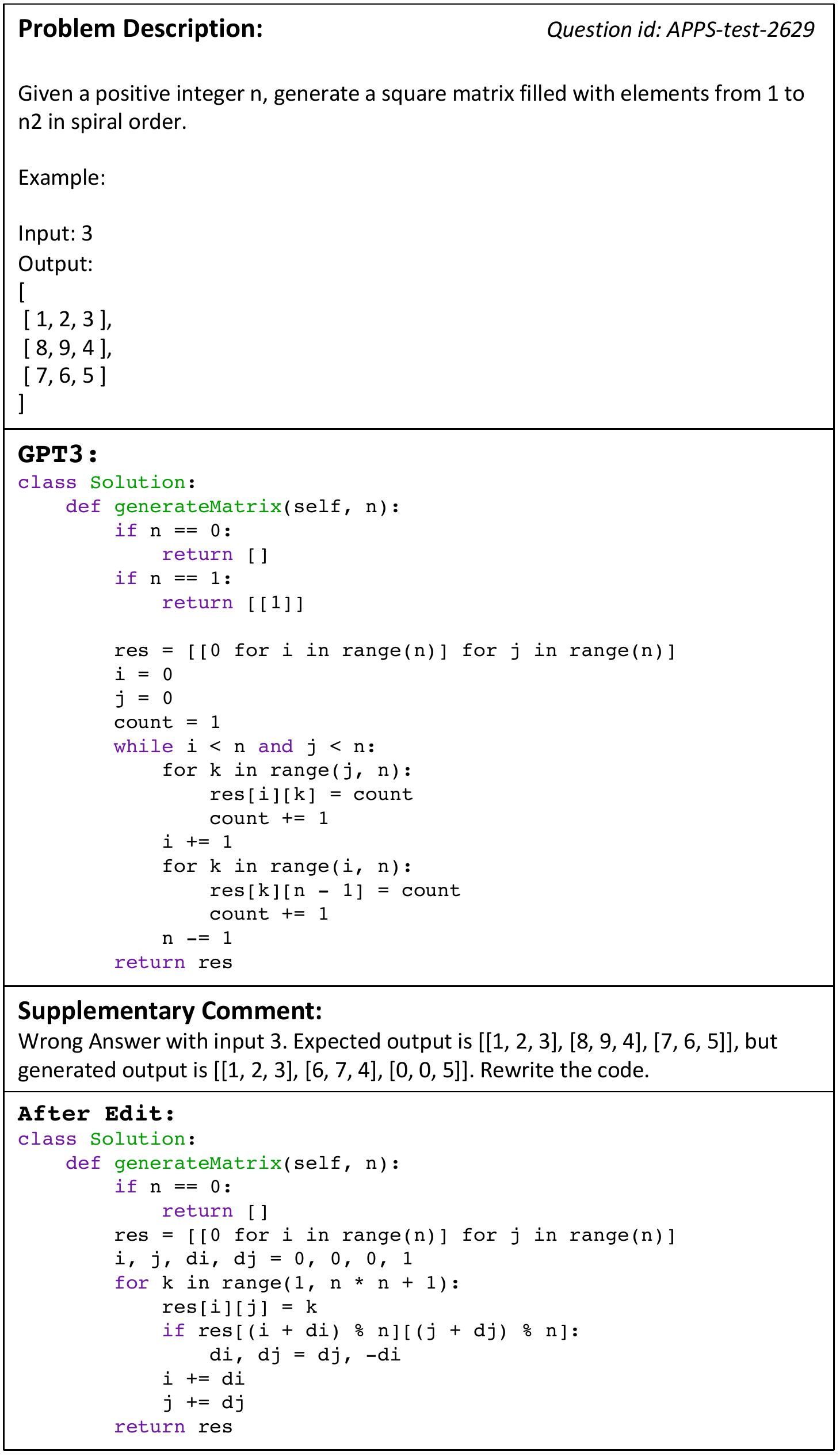}  
}
\subfigure[\scriptsize]{
\includegraphics[width=\columnwidth]{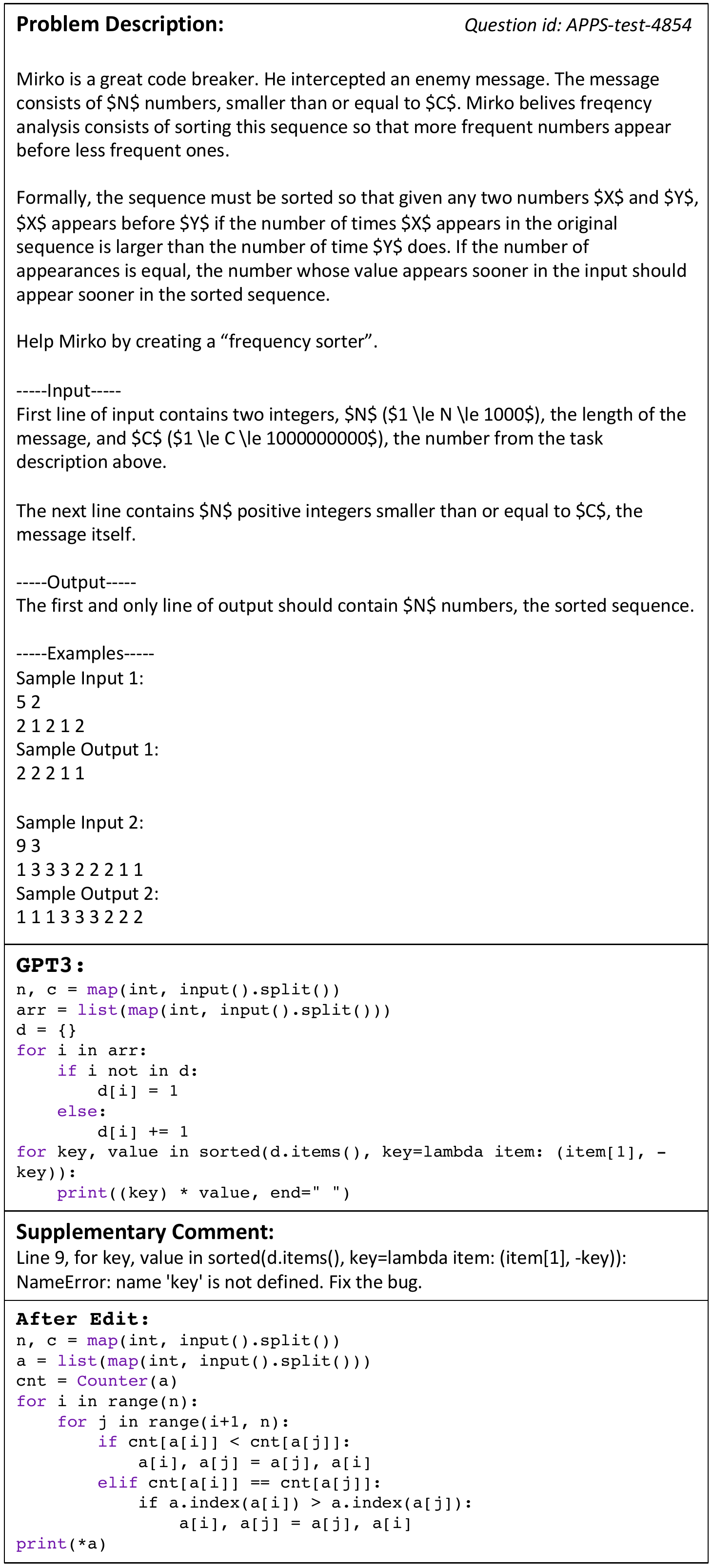}  
}
\caption{Case Study on APPS-test dataset using \textit{GPT3} model.}
\label{fig:case_study_test}
\vspace{-10pt}
\end{figure*}

\begin{figure*}[t]
\subfigure[\scriptsize]{
\label{fig:case_study_humaneval1}
\includegraphics[width=\columnwidth]{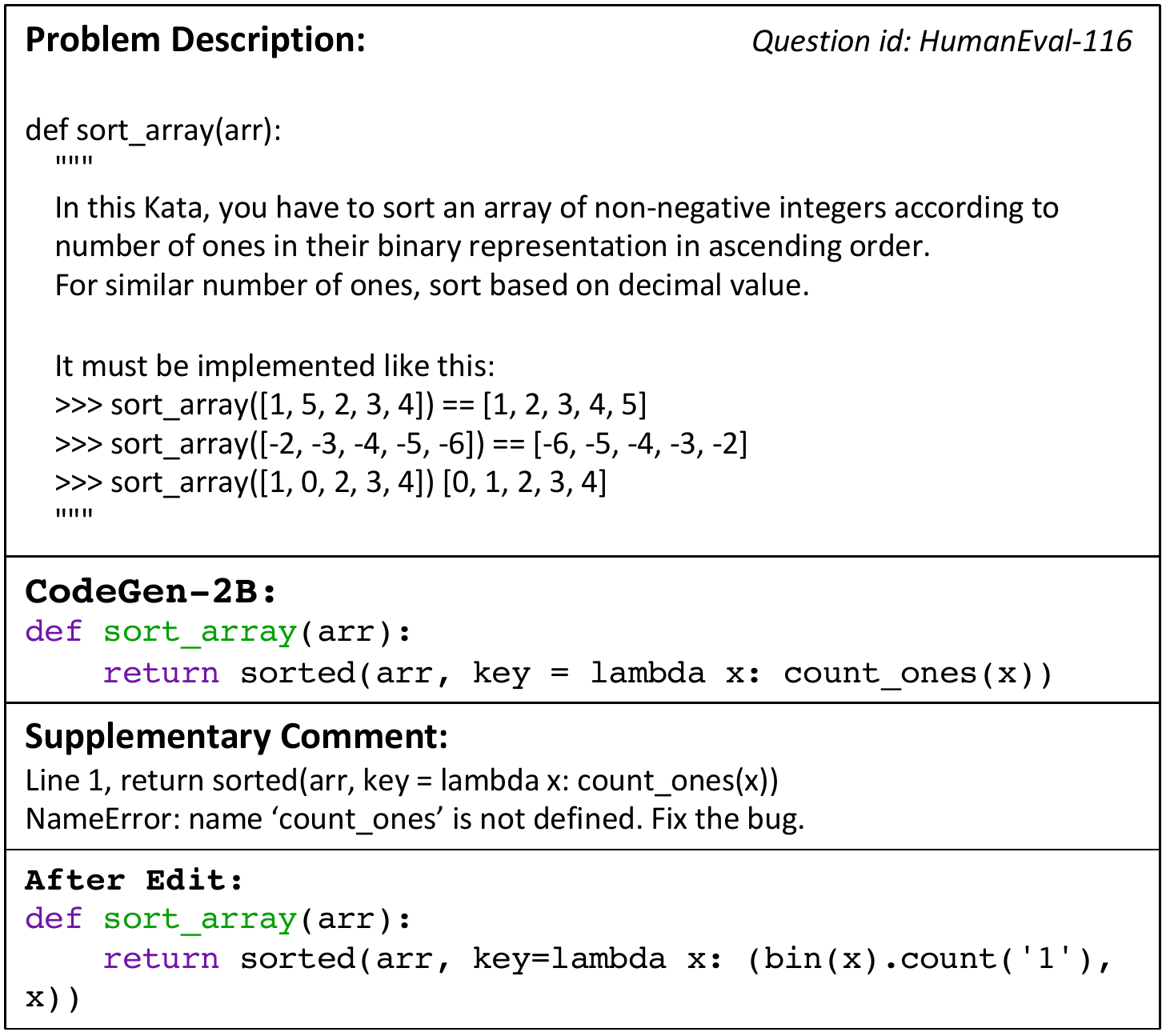}  
}
\subfigure[\scriptsize]{
\label{fig:case_study_humaneval2}
\includegraphics[width=\columnwidth]{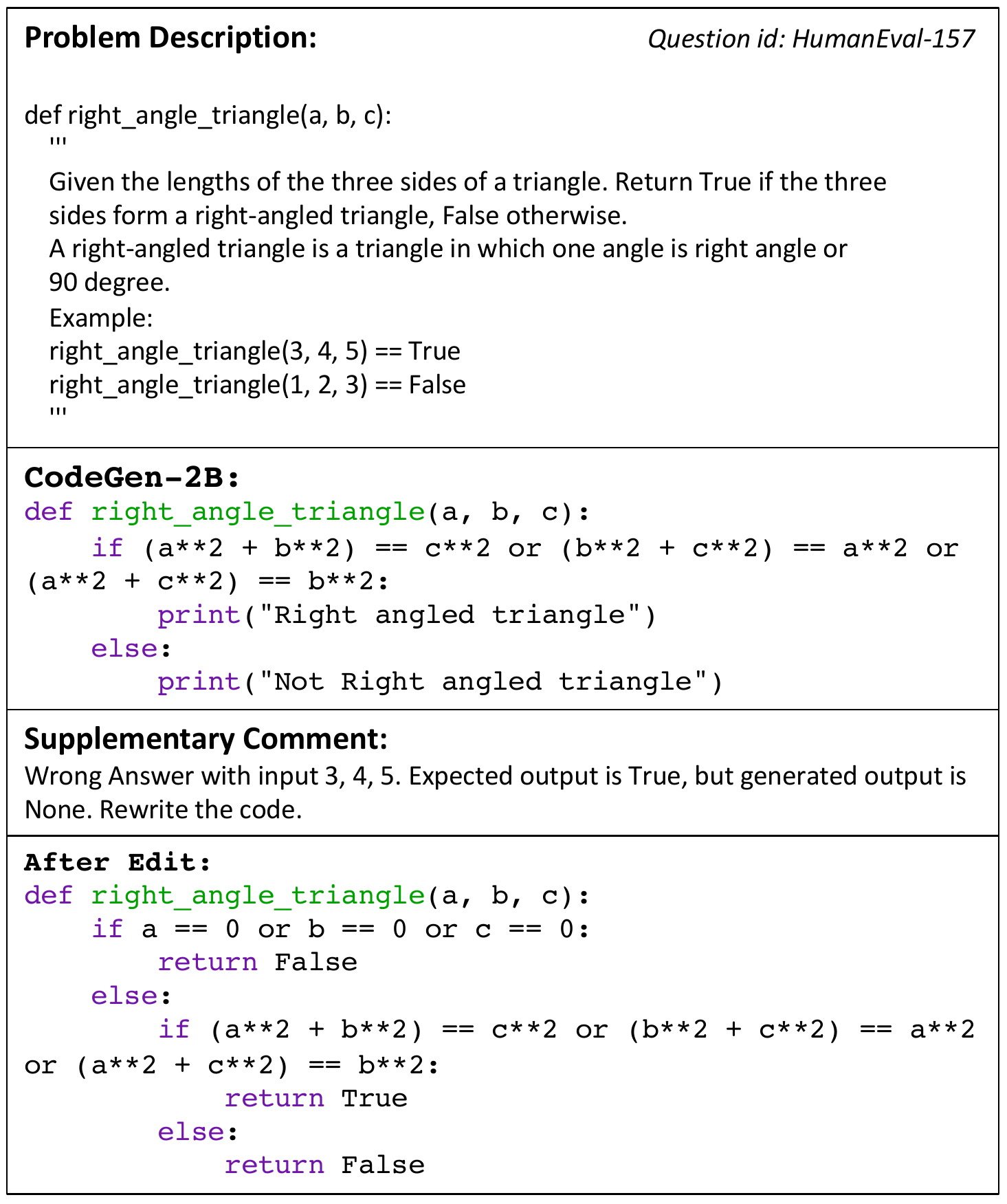}  
}
\caption{Case Study on HumanEval dataset using \textit{CodeGen-2B} model.}
\label{fig:case_study_humaneval}
\vspace{-10pt}
\end{figure*}

In Figure \ref{fig:case_study_dev} and \ref{fig:case_study_test} we show various programs generated by the \textit{GPT3}, its corresponding problem description (contains example test case) and the supplementary comment. Our fault-aware code editor concatenates these as input, and generate the edited code as the final output. We find that the edited code is simialr to the \textit{GPT3} output. In particular, the first few lines of the edited output are exactly the same as the output of \textit{GPT3}, and the subsequent code is also partially based on the content in \textit{GPT3} output. 
Through statistical analysis, we find that the common prefix between the two sequences accounted for 19.10\% of the edited output on the APPS-dev and APPS-test datasets. While this does not account for similarities in the intermediate content, it is sufficient evidence to demonstrate the impact of the LLM output on the edited code. As for the HumanEval benchmark, we also show case studies in Figure \ref{fig:case_study_humaneval}.

\end{document}